\documentclass[a4paper,fleqn]{cas-dc}

\usepackage[authoryear,longnamesfirst]{natbib}
\usepackage{etoolbox}
\makeatletter
\patchcmd{\algorithmic}{\itemsep\z@}{\itemsep=0.3pt}{}{}
\makeatother

\usepackage[section]{placeins}
\usepackage{algorithm}
\usepackage{algpseudocode}

\makeatletter
\renewcommand{\ALG@beginalgorithmic}{\scriptsize}
\makeatother
\algrenewcommand\algorithmicindent{1.0em}
\usepackage{enumitem}
\usepackage{float}
\usepackage{caption}
\usepackage{graphicx}
\usepackage{cuted}

\def\tsc#1{\csdef{#1}{\textsc{\lowercase{#1}}\xspace}}
\tsc{WGM}
\tsc{QE}

\begin{document}
\let\WriteBookmarks\relax
\def\floatpagepagefraction{1}
\def\textpagefraction{.001}

\shorttitle{}    

\shortauthors{}  

\title [mode = title]{Auditable Machine Unlearning for Privacy-Compliant Ransomware Detection Using Multi-Shard SISA and Deep Reinforcement Learning}
%

\author[1]{Jannatul Ferdous}[orcid=0000-0002-9612-0482]
\cormark[1]
\ead{jferdous@csu.edu.au}
\credit{Conceptualization, Methodology, Software, Writing - original draft}

\affiliation[1]{organization={School of Computing, Mathematics and Engineering, Charles Sturt University},
    city={Wagga Wagga},
    state={NSW},
    postcode={2650},
    country={Australia}}

\author[2]{Rafiqul Islam}
\credit{Supervision, Validation, Writing - review \& editing}

\affiliation[2]{organization={School of Computing, Mathematics and Engineering, Charles Sturt University},
    city={Albury},
    state={NSW},
    postcode={2640},
    country={Australia}}

\author[3,4]{Md Zahidul Islam}
\credit{Supervision, Writing - review \& editing}

\affiliation[3]{organization={School of Computing, Mathematics and Engineering, Charles Sturt University},
    addressline={Panorama Avenue},
    city={Bathurst},
    state={NSW},
    postcode={2795},
    country={Australia}}

\affiliation[4]{organization={AI and Cyber Futures Centre, Charles Sturt University},
    addressline={Panorama Avenue},
    city={Bathurst},
    state={NSW},
    postcode={2795},
    country={Australia}}

\cortext[1]{Corresponding author}


\begin{abstract}
Ransomware poses an escalating cybersecurity threat as attackers continuously modify behavioural patterns to evade static defenses. Although existing machine learning-based detectors often achieve strong predictive performance, they generally assume fixed training data and do not support the selective removal of previously learned samples. This limitation conflicts with privacy regulations such as the GDPR and CCPA, which require the removal of sensitive user data upon request. To address this challenge, we propose an auditable ransomware detection and unlearning framework that integrates deep reinforcement learning with multi-shard SISA retraining. In the proposed system, a Double Deep Q-Network (DDQN) learns a reward-guided detection policy from behavioral features under asymmetric security costs, while multi-shard SISA enables privacy-compliant selective sample removal through shard-level retraining. The framework was evaluated using four criteria: utility preservation, oracle-based forgetting validation, membership inference auditing, and computational efficiency. On a balanced Windows 11 behavioral dataset comprising 2,000 samples and 103 features, the baseline DDQN detector achieved an F1 score of 0.9925 and an AUC of 0.9983. The experimental results show that single-shard unlearning maintains minimal utility degradation and low oracle disagreement, whereas moderate shard counts ($M = 5$--$10$) provide the best efficiency-performance trade-off, reducing retraining time to 5--30~s compared with 80--330~s for full retraining. In addition, the membership inference scores remain close to 0.5 across most configurations, indicating limited privacy leakage after unlearning. These findings demonstrate that a privacy-compliant ransomware detection framework can jointly achieve high detection performance, auditable deletion verification, and efficient sample removal.
\end{abstract}




\maketitle

\section{Introduction}

Ransomware remains one of the most severe cybersecurity threats, with projected global damages exceeding \$265 billion annually by 2030~\cite{ref1}. By encrypting user data and demanding payment for decryption, ransomware attacks have caused substantial financial and operational damage across industries including healthcare, finance, and government services~\cite{ref2}. As ransomware campaigns continue to evolve through obfuscation, packing techniques, and rapidly changing variants, traditional signature-based detection mechanisms are becoming increasingly ineffective.

Behavior-based detection approaches have been widely explored to overcome these limitations. Instead of relying on static signatures, these methods analyze dynamic behavioral traces collected during program execution, including file system activity, registry modifications, process behavior, and network interactions of the program. Machine learning (ML) and deep learning (DL) techniques have demonstrated strong capabilities in identifying ransomware patterns from such behavioral telemetry~\cite{ref3,ref4}. Recently, reinforcement learning (RL) has attracted growing attention in cybersecurity because it surpasses the limitations of static classification rules; rather than learning a fixed input-output mapping from labeled examples, it can learn reward-guided decision policies~\cite{ref5}. This property is particularly relevant to ransomware detection, where evolving behaviors reduce the effectiveness of fixed-rule detectors and false negatives are substantially more costly than false positives. The RL reward structure can directly encode this asymmetric security cost by imposing stronger penalties for missed ransomware detections than for false alarms. This enables the detector to prioritize security-critical outcomes more directly than conventional symmetric classification objectives.

Despite advances in detection accuracy, an important challenge remains largely unaddressed: the ability to remove specific training samples from deployed machine learning models. Modern regulatory frameworks such as the General Data Protection Regulation (GDPR)~\cite{ref6} and California Consumer Privacy Act (CCPA)~\cite{ref7} impose strict requirements on data privacy management, including the ``right to be forgotten.'' These regulations require organizations to remove specific user data from trained models upon request. In cybersecurity datasets, such requirements may arise when sensitive traces are removed, mislabeled samples are discovered, or adversarial data are eliminated. Conventional retraining requires rebuilding the entire model after each deletion request, which is computationally expensive for continuously updated systems. Consequently, machine unlearning has emerged as an important research direction for removing the influence of selected training samples while preserving model utility and computational efficiency~\cite{ref8}.

Among existing approaches, the Sharded, Isolated, Sliced, and Aggregated (SISA) training paradigm proposed by Bourtoule \textit{et al.} provides an efficient strategy for scalable data removal~\cite{ref9,ref10}. In SISA, the training dataset is partitioned into independent shards and separate models are trained on each shard. When deletion requests occur, only the models corresponding to the affected shards require retraining, thereby significantly reducing the computational cost compared to full retraining~\cite{ref11}. Although SISA has been effective in supervised learning settings~\cite{ref9,ref10,ref12,ref13}, its application to RL-based cybersecurity systems remains largely unexplored. Unlike traditional classifiers, which learn a fixed mapping between inputs and labels, RL learns decision policies through reward-driven interactions with the environment. This allows detection models to learn reward-guided policies from behavioral state representations while explicitly encoding asymmetric security costs~\cite{ref14,ref15,ref16}. However, enabling machine unlearning in RL-based detection systems introduces additional challenges related to policy stability, verification of deletion accuracy, and privacy leakage.

To address these challenges, this study proposes an auditable machine unlearning framework for ransomware detection that integrates a Double Deep Q-Network (DDQN) agent with multi-shard SISA training. The framework uses DDQN to learn a reward-guided detection policy from behavioral ransomware features under asymmetric security costs, whereas SISA enables efficient shard-level retraining when deletion requests occur. To verify that deleted samples no longer influence the trained model, we incorporate oracle-based forgetting validation, in which the selectively unlearned model is compared with a reference model retrained from scratch on retained data only. To further assess post-unlearning privacy behavior, we evaluate membership-inference risk using Q-margin-based auditing. The proposed framework is evaluated using a behavioral ransomware dataset comprising 2{,}000 Windows~11 samples represented by 103 features extracted from sandbox execution traces. To strengthen the empirical study despite the moderate dataset size, we conduct five-fold cross-validation and evaluate multiple deletion scenarios spanning single-shard and multi-shard retraining, different shard granularities, repeated deletion rounds, and varying forget fractions. The evaluation focuses on four aspects: (i) utility preservation after unlearning, (ii) computational efficiency of shard-level retraining, (iii) deletion accuracy through oracle comparison, and (iv) privacy risk through membership-inference analysis. A preliminary version of this work appeared in~\cite{ref17} under a limited one-shard setting. The present study substantially extends that version by introducing multi-shard unlearning, oracle-based forgetting validation, membership-inference auditing, and a broader deletion-stress evaluation.

The main contributions of this study are as follows:
\begin{itemize}
    \item \textbf{DDQN--SISA framework for privacy-compliant ransomware unlearning:} We propose a privacy-compliant ransomware detection and unlearning framework that integrates a DDQN-based RL detector with multi-shard SISA. The DDQN learns a reward-guided detection policy under asymmetric security costs, whereas SISA enables selective shard-level retraining after deletion requests, avoiding full model retraining and supporting privacy-compliant data removal.

    \item \textbf{Auditable unlearning via dual-mechanism verification:} We introduce a dual-audit protocol that combines (a) oracle-verified forgetting, in which the unlearned model is compared with a reference model retrained exclusively on retained data to confirm that deleted samples no longer influence detection decisions, and (b) membership-inference privacy auditing based on ensemble Q-value margins to verify that deleted samples are indistinguishable from never-observed data. This dual-channel verification supports compliance-oriented auditing of deployed information security systems.

    \item \textbf{Systematic deletion-stress evaluation:} We conduct a comprehensive empirical study across 270 configurations spanning multiple shard counts ($M = 5, 10, 20$), deletion scopes (single- versus multi-shard), forget fractions (1\%, 5\%, and 10\%), and repeated deletion rounds (1, 5, and 10) to characterize the utility--privacy--efficiency trade-off and identify stable operating regimes for practical deployment.
\end{itemize}

The remainder of this paper is organized as follows. Section~2 reviews related work on ransomware detection, reinforcement learning in cybersecurity, and machine unlearning. Section~3 presents the proposed DDQN--SISA framework and verification protocols. Section~4 describes the experimental setup. Section~5 reports the empirical results. Section~6 discusses the implications and limitations of the study. Section~7 concludes the paper.

\section{Related Work}

This section reviews prior research in three interconnected areas: behavior-based ransomware detection, reinforcement learning in cybersecurity, and machine unlearning using SISA.

\subsection{Behavior-Based Ransomware Detection}

Behavior-based ransomware detection has emerged as an effective alternative to static signature-based approaches, which are vulnerable to obfuscation and polymorphic malware variants~\cite{ref18}. These methods detect ransomware by analyzing dynamic execution behavior, such as file system changes, registry activity, application programming interface (API) calls, and network communication patterns~\cite{ref19,ref20,ref21}.

Recent research has applied machine learning and deep neural networks to sandbox-generated behavioral traces to improve detection accuracy and generalization across ransomware families~\cite{ref18,ref22,ref21,ref23}. For example, swarm intelligence-enhanced frameworks utilize memory forensics features and feature-selection techniques to improve classification performance~\cite{ref24}, whereas zero-day detection approaches employ generative models, such as conditional variational autoencoders, to capture latent behavioral representations of unseen attacks~\cite{ref25}. In addition, explainable ransomware detection based on dynamic analysis has been explored in systems such as XRGuard and XRan to improve detection transparency and robustness~\cite{ref23,ref26}.

Although these approaches achieve high detection accuracy, they do not address the regulatory requirements for selective data deletion or model updates under privacy constraints. This limitation motivates the development of ransomware detection frameworks that combine strong detection capabilities with scalable machine unlearning mechanisms.

\subsection{Reinforcement Learning in Cybersecurity}

Deep reinforcement learning (DRL) has gained increasing attention in cybersecurity because it can learn reward-guided decision policies rather than fixed input-output mappings~\cite{ref27}. This property is particularly relevant for security tasks, where classification errors carry asymmetric operational costs. Value-based methods such as Deep Q-Networks (DQN) and Double Deep Q-Networks (DDQN) estimate state-action value functions to support stable and sample-efficient policy learning, while DDQN mitigates Q-value overestimation by decoupling action selection through the online network from action evaluation through the target network~\cite{ref28}. The resulting Q-value margin, defined as the difference \(Q(s,1)-Q(s,0)\) for a given behavioral state, provides a natural continuous confidence measure that can be used for ROC analysis and membership-inference auditing without additional calibration. However, existing DRL-based ransomware detection studies primarily focus on detection performance~\cite{ref29} or defense-oriented policy design~\cite{ref30} and do not address how learned Q-policies should be updated when training samples must be selectively removed under privacy regulations. Consequently, the effects of deletion requests on Q-value stability, policy behavior, and downstream privacy leakage in RL-based cybersecurity systems remain insufficiently explored.

\subsection{Machine Unlearning and SISA Framework}

Machine unlearning aims to remove the influence of specific training samples from trained models~\cite{ref31,ref32}. While early approaches often relied on full retraining, the SISA framework reduces recomputation by restricting retraining to the affected data shards~\cite{ref9,ref31}. Prior SISA studies have primarily focused on supervised learning benchmarks and reported favorable efficiency and privacy-related trends~\cite{ref32,ref33}. However, the application of shard-based retraining to RL agents, particularly in behavior-based security systems, remains largely unexplored. Furthermore, existing unlearning evaluations often emphasize runtime reduction while providing limited empirical mechanisms for verifying deletion accuracy or auditing privacy leakage under adversarial conditions.

\subsection{Research Gap}

Although significant advances have been made in ransomware detection, RL for cybersecurity, and scalable machine unlearning, their integration has not been systematically investigated in ransomware detection. In particular, the combination of multi-shard SISA retraining with value-based deep reinforcement learning, oracle-based forgetting verification, and membership-inference privacy auditing has not been thoroughly studied in this setting. This study addresses this gap by proposing an auditable DDQN--SISA unlearning framework specifically designed for behavior-based ransomware detection.

\section{Overall System Architecture}

\begin{figure*}[!t]
    \centering
    \vspace{-6pt}
    \includegraphics[width=\textwidth]{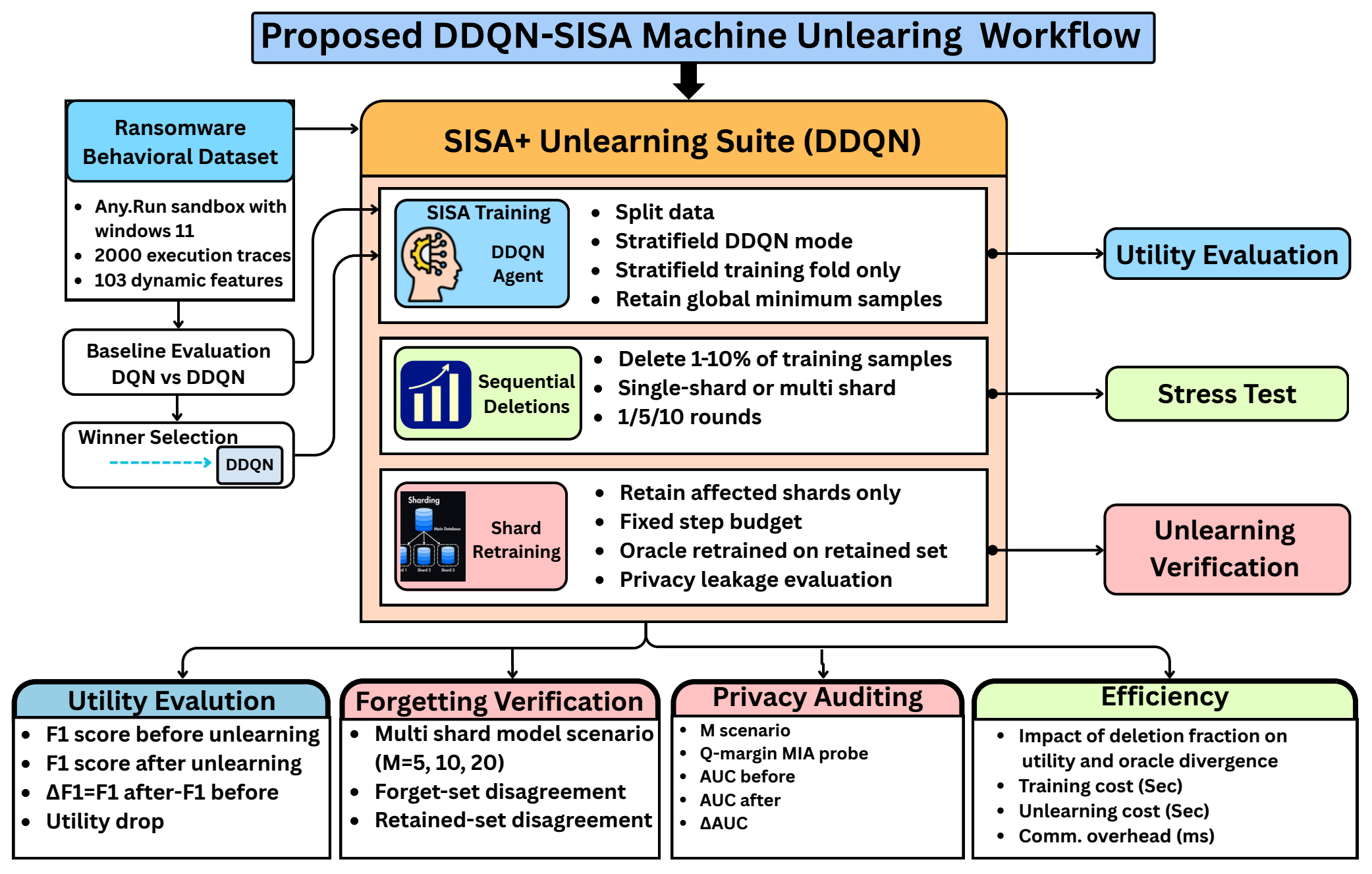}
    \vspace{-4pt}
    \caption{Proposed DDQN--SISA machine unlearning workflow. The framework integrates behavioral dataset construction, RL-based baseline model selection, multi-shard SISA retraining, and audit-oriented evaluation for privacy-compliant ransomware detection.}
    \vspace{-6pt}
    \label{fig:overall_workflow}
\end{figure*}

The proposed architecture establishes a verifiable framework for privacy-compliant ransomware detection by integrating behavioral dataset representation, value-based DRL, multi-shard SISA unlearning, and verification mechanisms. As illustrated in Fig.~\ref{fig:overall_workflow}, the workflow begins with the construction of a balanced Windows~11 behavioral dataset (Section~\ref{sec:dataset_representation}) and the definition of the unlearning objective (Section~\ref{sec:problem_formulation}). Ransomware detection is then formulated as a Markov decision process (Section~\ref{sec:rl_environment}), and DQN/DDQN agents are trained under stratified cross-validation to select the best model (Section~\ref{sec:baseline_rl_detector}). The selected DDQN agent is embedded in a multi-shard SISA framework ($M=5,10,$ and $20$), enabling stratified training, sequential deletion, and shard retraining (Section~\ref{sec:sisa_framework}). Finally, a comprehensive evaluation framework (Section~\ref{sec:unlearning_evaluation}) assesses utility preservation, oracle-verified forgetting, privacy risk via Q-margins, and computational efficiency. Together, these components form a unified system in which detection accuracy, deletion fidelity, privacy auditing, and efficiency are jointly optimized under sequential deletion conditions.

\subsection{Dataset Representation}
\label{sec:dataset_representation}

To evaluate privacy-aware ransomware detection under realistic operating conditions, we constructed a balanced behavioral dataset comprising 2{,}000 Windows executable samples, including 1{,}000 ransomware and 1{,}000 benign programs. The dataset was designed to reflect contemporary Windows~11 environments and overcome limitations commonly observed in legacy Cuckoo-based corpora.

\paragraph{Sample Collection.}
Ransomware samples were collected from MalwareBazaar~\cite{ref36} and VirusShare~\cite{ref37} from 30 distinct ransomware families. We used the following two criteria to select ransomware families:
\begin{enumerate}
    \item Threat-intelligence reports identified the family as highly frequent and globally impactful in 2019~\cite{ref38}, 2020~\cite{ref39}, 2021~\cite{ref40}, 2022~\cite{ref41}, 2023~\cite{ref42}, and 2024~\cite{ref43}.
    \item Confirmation was available from at least two independent reports by well-regarded cybersecurity firms.
\end{enumerate}

Using this selection strategy, prominent families such as LockBit, Conti, BlackCat, MedusaLocker, Phobos, and WannaCry were included to capture diverse encryption strategies, persistence mechanisms, and command-and-control behaviors.

Individual ransomware samples were selected following the methodology outlined in~\cite{ref44}, employing three criteria based on VirusTotal vendor detections. A sample was retained if it satisfied three conditions: (i) detection as malicious by at least 45 antivirus engines; (ii) explicit identification as ransomware by at least 15 engines; and (iii) majority agreement on ransomware family classification. This multi-vendor validation strengthens labeling confidence and dataset integrity. Table~\ref{tab:ransomware_families} shows the ransomware-family distribution and sample counts.
Benign samples were collected from trusted software repositories, including SnapFiles~\cite{ref45}, PortableApps.com~\cite{ref46}, and GitHub~\cite{ref47}, to ensure diversity across utility tools, productivity software, and system applications.

\begin{table}[t]
\caption{Ransomware families included in this study with sample counts.}
\label{tab:ransomware_families}
\centering
\scriptsize
\setlength{\tabcolsep}{4pt}
\renewcommand{\arraystretch}{0.85}
\begin{tabular}{ll|ll}
\hline
\textbf{Family} & \textbf{No.} & \textbf{Family} & \textbf{No.} \\
\hline
LockBit             & 114 & Phobos           & 22 \\
GandCrab (Grancrab) & 97  & WannaCry         & 21 \\
MedusaLocker        & 63  & WastedLocker     & 20 \\
NetWalker           & 56  & BlackMatter      & 20 \\
Conti               & 55  & BlackBasta       & 20 \\
Babuk               & 52  & Ryuk             & 20 \\
Cerber              & 49  & RagnarLocker     & 19 \\
Maze                & 47  & Mespinoza (Pysa) & 19 \\
Sodinokibi / REvil  & 43  & AvosLocker       & 15 \\
DarkSide            & 41  & Avadon           & 14 \\
Dharman             & 36  & MountLocker      & 11 \\
Thanos              & 31  & Exorcist         & 11 \\
Teslacrypt          & 31  & Mallox           & 10 \\
Makop               & 23  & Nefilim          & 10 \\
Akira               & 22  & BlueSky          & 8  \\
\hline
\multicolumn{2}{l|}{\textbf{Total ransomware families = 30}} &
\multicolumn{2}{l}{\textbf{Total ransomware samples = 1000}} \\
\hline
\end{tabular}
\end{table}

\paragraph{Sample Execution.}
All samples were executed in a Windows~11 environment using the ANY.RUN sandbox~\cite{ref48}, which was selected for its support for modern operating systems, rich behavioral telemetry, and improved resistance to sandbox-evasion techniques compared with legacy Cuckoo-based systems~\cite{ref49}. Each execution generated a detailed JSON report containing approximately 250 behavioral fields.

\paragraph{Preprocessing.}
To ensure reproducibility and prevent data leakage, preprocessing steps, including biased entry removal, feature selection, numeric encoding, and cross-validation-aware standardization, produced an optimized dataset covering file system, registry, process, API, network, CryptoAPI, incident-rule, reputation, and multi-process behaviors. The final output was a 103-dimensional feature vector suitable for ML- and RL-based ransomware detection.

The resulting dataset provides a balanced class distribution, family diversity, modern OS fidelity, and strong labeling reliability, making it suitable for evaluating detection accuracy, privacy leakage, and machine unlearning performance under realistic threat scenarios.

\begingroup
\setlength{\abovedisplayskip}{3pt}
\setlength{\belowdisplayskip}{3pt}
\setlength{\abovedisplayshortskip}{1.5pt}
\setlength{\belowdisplayshortskip}{1.5pt}

\subsection{Problem Formulation and Unlearning Objective}
\label{sec:problem_formulation}

Let the full behavioral ransomware dataset be defined as
\begin{equation}
\mathcal{D}=\{(\mathbf{x}_i, y_i)\}_{i=1}^{N}
\label{eq:dataset}
\end{equation}
where \(\mathbf{x}_i \in \mathbb{R}^{103}\) denotes a 103-dimensional behavioral feature vector and \(y_i \in \{0,1\}\) represents the class label. Specifically, \(y_i=1\) indicates ransomware and \(y_i=0\) denotes benign behavior. The dataset is balanced with \(N=2000\).

The learning objective is to construct a classifier
\begin{equation}
f_{\theta}:\mathbb{R}^{103}\rightarrow \{0,1\}
\label{eq:classifier}
\end{equation}
that maximizes detection performance while allowing the removal of a subset of training samples,
\begin{equation}
\mathcal{D}_f \subset \mathcal{D}
\label{eq:forgetset}
\end{equation}
without retraining on the entire dataset.

Let the retained dataset after deletion be
\begin{equation}
\mathcal{D}_r=\mathcal{D}\setminus\mathcal{D}_f
\label{eq:retainedset}
\end{equation}

After deletion, the updated model should approximate the oracle model trained exclusively on the retained data:
\begin{equation}
f_{\theta}^{(-\mathcal{D}_f)} \approx f_{\theta_{\mathrm{oracle}}}
\label{eq:oracle_target}
\end{equation}
where
\[
\theta_{\mathrm{oracle}}=\mathrm{Train}(\mathcal{D}_r)
\]

The objective is therefore to obtain an updated model \(f_{\theta}^{(-\mathcal{D}_f)}\) satisfying:
\begin{itemize}
    \item utility preservation,
    \item correctness of unlearning through oracle alignment, and
    \item reduced membership distinguishability.
\end{itemize}

Thus, machine unlearning is formulated as a constrained optimization problem that balances detection performance, deletion fidelity, privacy risk, and computational cost.

\subsection{Design of the Reinforcement Learning Environment}
\label{sec:rl_environment}

Ransomware detection was formulated as a Markov decision process (MDP),
\[
\mathcal{M}=(S,A,P,R,\gamma)
\]
where the state space, action space, transition dynamics, reward function, and discount factor are defined as follows.

\paragraph{State space \(S\).}
Each behavioral feature vector corresponds to a state:
\begin{equation}
s_t=\mathbf{x}_i, \qquad \mathbf{x}_i \in \mathbb{R}^{103}
\label{eq:state}
\end{equation}

\paragraph{Action space \(A\).}
The agent classifies each sample as benign or ransomware. We define the binary action space as
\begin{equation}
A=\{0,1\}
\label{eq:action}
\end{equation}
where \(0\) and \(1\) denote benign and ransomware samples, respectively.

\paragraph{Transition dynamics \(P\).}
Transitions are deterministic and follow the sequential order of the dataset samples. After processing sample \(\mathbf{x}_i\), the next state becomes
\begin{equation}
s_{t+1}=\mathbf{x}_{(i+1)\bmod N}
\label{eq:transition}
\end{equation}
The modulo operator ensures that after the final sample, the sequence returns to the beginning.

Because transitions are independent of actions and follow a fixed order, the environment reduces to a contextual decision process with immediate reward feedback.

\paragraph{Reward function R.}
The asymmetric reward function reflects the unequal security impact of classification errors in ransomware detection. In practice, false negatives can cause severe damage through data encryption and operational disruption, whereas false positives mainly result in alerts or inspections. Therefore, missed ransomware detections are penalized more strongly than false positives. The reward is defined as
\begin{equation}
R(y_t,a_t)=
\begin{cases}
+1, & a_t=y_t,\\
-2, & y_t=1,\ a_t=0,\\
-0.5, & y_t=0,\ a_t=1.
\end{cases}
\label{eq:reward}
\end{equation}
This formulation encourages the agent to prioritize security-critical outcomes.

\paragraph{Discount factor \(\gamma\).}
We set \(\gamma=0.1\), emphasizing immediate rewards over long-term returns. In ransomware detection, each classification decision has an independent value without delayed consequences. The low discount factor therefore discourages long-horizon dependencies and focuses the learning process on immediate classification accuracy.

\begingroup
\setlength{\abovedisplayskip}{3pt}
\setlength{\belowdisplayskip}{3pt}
\setlength{\abovedisplayshortskip}{1.5pt}
\setlength{\belowdisplayshortskip}{1.5pt}

\subsection{Baseline RL Detector}
\label{sec:baseline_rl_detector}

This study adopts value-based DRL rather than policy-gradient or actor-critic approaches for three task-specific reasons. First, ransomware detection is a discrete binary classification problem, namely benign versus ransomware, for which direct Q-value estimation over a two-action space is simpler, more stable, and more sample-efficient than policy-gradient methods that optimize a stochastic policy over a continuous action distribution. Second, value-based methods produce explicit action-value margins \(Q(s,1)-Q(s,0)\), which serve a dual purpose beyond detection: they provide continuous confidence scores for ROC analysis and oracle comparison and act as membership-inference audit signals, a representation that policy-gradient methods do not naturally produce. Third, off-policy learning via experience replay decouples data collection from policy updates, improving sample efficiency and integrating naturally with the shard-isolated SISA architecture, where each shard independently estimates \(Q(s,a)\) without requiring coordinated policy updates across shards, a property that on-policy actor-critic methods cannot easily satisfy.

Two value-based algorithms were evaluated using 5-fold stratified cross-validation under fixed training step budgets:
\begin{itemize}
    \item Deep Q-Network (DQN)
    \item Double Deep Q-Network (DDQN)
\end{itemize}

Both models were implemented with identical configurations, differing only in TD-target computation. They use the same network architecture, replay-buffer capacity, learning rate, discount factor, exploration schedule, batch size, target-network updates, and random seeds. Consequently, any performance difference arises from the TD target formulation rather than confounding hyperparameters.

\paragraph{Temporal-Difference Optimization.}
A parameterized Q-network \(Q_{\theta}(s,a)\) estimates state--action values. Experiences \((s_t,a_t,r_t,s_{t+1},d_t)\) are stored in a replay buffer \(B\) and sampled uniformly. Both DQN and DDQN update the network by minimizing the TD loss,
\begin{equation}
\mathcal{L}(\theta)=
\mathbb{E}_{(s_t,a_t,r_t,s_{t+1},d_t)\sim B}
\left[
\ell\bigl(Q_{\theta}(s_t,a_t)-\mathcal{Y}_t\bigr)
\right]
\label{eq:td_loss}
\end{equation}
where \(\ell(\cdot)\) is the Huber (smooth L1) loss and \(\mathcal{Y}_t\) is the TD target.

\paragraph{DQN Target.}
In DQN, the TD target is computed as
\begin{equation}
\mathcal{Y}_t^{\mathrm{DQN}}
=
r_t+\gamma(1-d_t)\max_{a'}Q_{\theta^-}(s_{t+1},a')
\label{eq:dqn_target}
\end{equation}
The same value function is used for both action selection and evaluation, which may lead to overestimation bias.

\paragraph{DDQN Target.}
In DDQN, overestimation is mitigated by decoupling action selection and evaluation:
\begin{equation}
\mathcal{Y}_t^{\mathrm{DDQN}}
=
r_t+\gamma(1-d_t)\,
Q_{\theta^-}\!\left(
s_{t+1},
\arg\max_{a'}Q_{\theta}(s_{t+1},a')
\right)
\label{eq:ddqn_target}
\end{equation}
Here, \(Q_{\theta}\) (online network) selects the action and \(Q_{\theta^-}\) (target network) evaluates it. This decoupling matches the implementation used in the codebase and reduces value overestimation while preserving computational efficiency.
\endgroup

\subsection{SISA Unlearning Framework for Deep Reinforcement Learning}
\label{sec:sisa_framework}

Machine learning removes the influence of specific training samples from a trained model without retraining on the entire data set. Based on achieving the highest mean F1-score across folds, DDQN was selected as the winning model and integrated into a multi-shard SISA architecture. This framework incorporates four core components: Sharding, Isolation, Slicing, and Aggregation to enable machine unlearning for DDQN-based ransomware detection by partitioning data into shards, training separate models per shard, and aggregating predictions. The overall workflow of the proposed multi-shard SISA framework is illustrated in Fig.~\ref{fig:multishard_sisa_framework}. This decomposition localizes the influence of the sample within the shard-level models, enabling selective retraining under deletion requests. The design is architecturally compatible with DDQN because each shard model maintains an independent Q-network and a replay buffer. Consequently, retraining one shard after a deletion request modifies only that shard's Q-value policy without perturbing the value functions of other shards, thereby facilitating localized and auditable unlearning relative to globally trained models.

\begin{figure*}[!t]
    \centering
    \vspace{-6pt}
    \includegraphics[width=\textwidth]{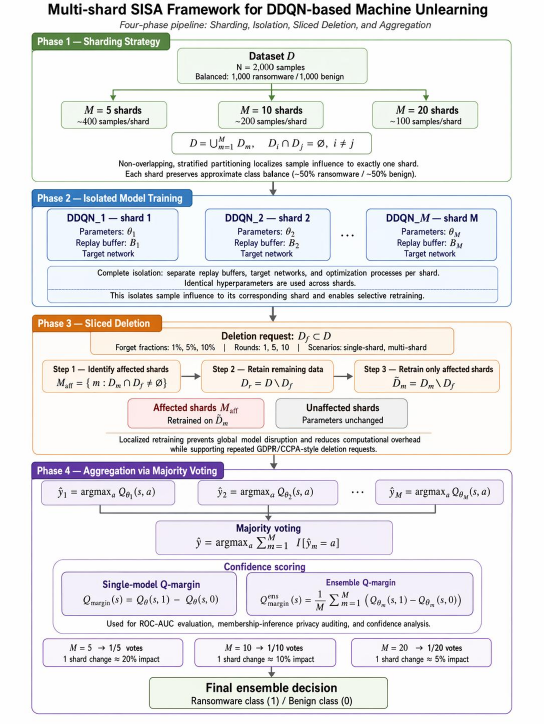}
    \vspace{-3pt}
    \caption{Multi-shard SISA framework for DDQN-based machine unlearning, showing four phases: stratified sharding, isolated shard-wise DDQN training, localized deletion-driven retraining of affected shards only, and majority-vote aggregation with ensemble Q-margin scoring.}
    \vspace{-4pt}
    \label{fig:multishard_sisa_framework}
\end{figure*}

\begingroup
\setlength{\abovedisplayskip}{3pt}
\setlength{\belowdisplayskip}{3pt}
\setlength{\abovedisplayshortskip}{1.5pt}
\setlength{\belowdisplayshortskip}{1.5pt}

\subsubsection{Sharding Strategy (Phase 1)}
\label{subsubsec:sharding}

The framework begins by partitioning the full dataset \(\mathcal{D}\) (\(N=2{,}000\) samples) into \(M\) non-overlapping shards. Formally,
\begin{equation}
\mathcal{D}=\bigcup_{m=1}^{M}\mathcal{D}_m,
\qquad
\mathcal{D}_i \cap \mathcal{D}_j = \emptyset,\; i \neq j
\label{eq:sharding}
\end{equation}

In our experiments, \(M \in \{5,10,20\}\). Shards are approximately equal in size and preserve class balance through stratified allocation. Three sharding configurations were evaluated:
\vspace{-2pt}
\begin{itemize}
    \setlength{\itemsep}{1pt}
    \setlength{\parskip}{0pt}
    \setlength{\parsep}{0pt}
    \setlength{\topsep}{2pt}
    \item \(M=5\): approximately 400 samples per shard (coarse granularity),
    \item \(M=10\): approximately 200 samples per shard (optimal balance),
    \item \(M=20\): approximately 100 samples per shard (fine granularity).
\end{itemize}
\vspace{-2pt}
Each shard maintains approximately equal distributions of ransomware (50\%) and benign (50\%) samples, ensuring that each DDQN model receives representative training data for the binary classification task. This stratification ensures every shard contains a balanced mix of malware behaviors, preventing any shard from being biased toward one class. Non-overlapping partitioning keeps the sample influence localized to one shard. This property enables the framework to identify which shards are "affected" by a deletion request, making selective retraining possible.
\endgroup

\begingroup
\setlength{\abovedisplayskip}{3pt}
\setlength{\belowdisplayskip}{3pt}
\setlength{\abovedisplayshortskip}{1.5pt}
\setlength{\belowdisplayshortskip}{1.5pt}

\subsubsection{Isolated Model Training (Phase 2)}
\label{subsubsec:isolation}

After sharding, independent DDQN models were trained on each shard with complete isolation, including separate replay buffers, target networks, and identical hyperparameters. This isolation ensures that the influence of any training sample remains localized to its corresponding shard, thereby enabling efficient selective retraining. In the implementation, each shard model is instantiated and trained independently using the same step budget and optimization settings.

\subsubsection{Sliced Deletion (Phase 3)}
\label{subsubsec:slicing}

Slicing refers to the controlled removal of selected training samples to simulate deletion requests. When deletion requests are received, the framework identifies and removes samples through a systematic deletion process.

\paragraph{Step 1: Identify affected shards.}
Only shards that contain at least one sample from the deletion set \(\mathcal{D}_f\) require retraining. The affected shards are identified as
\begin{equation}
\mathcal{M}_{\mathrm{aff}}
=
\left\{
m : \mathcal{D}_m \cap \mathcal{D}_f \neq \emptyset
\right\}
\label{eq:affected_shards}
\end{equation}

\paragraph{Step 2: Compute retained dataset.}
The retained dataset contains all training samples except those explicitly deleted and serves as the ground truth for oracle verification. Let the deletion subset satisfy \(\mathcal{D}_f \subset \mathcal{D}\). The retained dataset is then defined as
\begin{equation}
\mathcal{D}_r=\mathcal{D}\setminus\mathcal{D}_f
\label{eq:retained_dataset_phase3}
\end{equation}

\paragraph{Step 3: Retrain only affected shards.}
For each affected shard \(m \in \mathcal{M}_{\mathrm{aff}}\), the corresponding DDQN\(_m\) model is retrained on its retained subset
\begin{equation}
\widetilde{\mathcal{D}}_m=\mathcal{D}_m\setminus\mathcal{D}_f
\label{eq:retrained_shard}
\end{equation}

The unaffected shards remain unchanged. This localized retraining prevents global model disruption and substantially reduces computational overhead.

Deletion scenarios were evaluated under realistic compliance-oriented conditions:
\begin{itemize}[leftmargin=*, topsep=2pt, itemsep=1pt, parsep=0pt, partopsep=0pt]
    \item forget fractions: 1\%, 5\%, and 10\% (small, moderate, and aggressive deletions),
    \item deletion rounds: 1, 5, and 10 (cumulative deletion stress),
    \item scenario types: single-shard and multi-shard deletion patterns.
\end{itemize}
These combinations simulate GDPR/CCPA compliance requirements under repeated deletion requests over time.

\subsubsection{Aggregation via Majority Voting (Phase 4)}
\label{subsubsec:aggregation}

Individual shard predictions are combined through majority voting to obtain the final ensemble decision. For a given input state \(s\), each shard model produces a prediction
\begin{equation}
\hat{y}_m=\arg\max_{a \in A} Q_{\theta_m}(s,a)
\label{eq:shard_prediction}
\end{equation}

The final ensemble prediction is obtained via majority voting across shard outputs:
\begin{equation}
\hat{y}
=
\arg\max_{a \in A}
\sum_{m=1}^{M}\mathbb{I}[\hat{y}_m=a]
\label{eq:majority_vote}
\end{equation}
where \(\mathbb{I}[\cdot]\) is the indicator function that equals 1 if the condition holds and 0 otherwise.

For confidence-based analysis and privacy auditing, the framework computes Q-margins derived from the learned action-value function. For a single model, the Q-margin for state \(s\) is defined as
\begin{equation}
Q_{\mathrm{margin}}(s)=Q_{\theta}(s,1)-Q_{\theta}(s,0)
\label{eq:qmargin_single}
\end{equation}

This provides a continuous confidence score for ROC-AUC analysis.

In the multi-shard SISA setting, the ensemble Q-margin is computed by averaging shard-level margins:
\begin{equation}
Q_{\mathrm{margin}}^{\mathrm{ens}}(s)
=
\frac{1}{M}
\sum_{m=1}^{M}
\left(
Q_{\theta_m}(s,1)-Q_{\theta_m}(s,0)
\right)
\label{eq:qmargin_ensemble}
\end{equation}

This ensemble Q-margin represents the average confidence differential across all shard models and is used for ROC-AUC evaluation, membership-inference privacy auditing, and confidence-based score analysis. This aggregation mechanism ensures robustness because only the affected shards are updated during unlearning, whereas the unaffected shards retain their original parameters. Majority voting stabilizes predictions and preserves detection performance despite localized model updates.
\endgroup

\begingroup
\setlength{\abovedisplayskip}{3pt}
\setlength{\belowdisplayskip}{3pt}
\setlength{\abovedisplayshortskip}{1.5pt}
\setlength{\belowdisplayshortskip}{1.5pt}

\subsection{Unlearning Evaluation Framework}
\label{sec:unlearning_evaluation}

We evaluated the framework across four dimensions: utility preservation, oracle-aligned forgetting, membership-inference privacy risk, and computational efficiency. These dimensions ensure that deletion requests are implemented while maintaining performance, privacy, and manageable system overhead.

\subsubsection{Utility Preservation}
\label{subsubsec:utility_preservation}

Utility preservation measures the impact of unlearning on detection performance. Let \(F1_{\mathrm{before}}\) denote the F1-score of the model prior to deletion and \(F1_{\mathrm{after}}\) denote the F1-score after selective retraining. Utility variation is defined as
\begin{equation}
\Delta F1 = F1_{\mathrm{after}} - F1_{\mathrm{before}}
\label{eq:delta_f1}
\end{equation}

A small magnitude of \(\Delta F1\) indicates stable performance after unlearning. To explicitly quantify degradation, we report the utility drop as
\begin{equation}
U_{\mathrm{drop}}=\max\!\left(0,\,F1_{\mathrm{before}}-F1_{\mathrm{after}}\right)
\label{eq:utility_drop}
\end{equation}

This definition ensures a non-negative performance-loss measure. Consequently, the utility drop is zero whenever the post-unlearning performance improves or remains unchanged.

\subsubsection{Oracle-Verified Forgetting}
\label{subsubsec:oracle_forgetting}

Oracle verification assesses whether unlearning behaves as if deleted samples had never been used for training. The oracle model is retrained from scratch on the retained dataset as a gold-standard reference.

To perform oracle verification, a separate DDQN model was trained using only the retained training data,
\begin{equation}
\mathcal{D}_r=\mathcal{D}\setminus\mathcal{D}_f
\label{eq:oracle_retained}
\end{equation}
The predictions of the SISA-unlearned model were compared with those of the oracle model on:
\vspace{-4pt}
\begin{itemize}
    \setlength{\itemsep}{0pt}
    \setlength{\parskip}{0pt}
    \setlength{\parsep}{0pt}
    \setlength{\topsep}{1pt}
    \item the retained set, to verify that non-deleted data remain stable, and
    \item the forget set, to assess whether deleted samples no longer influence the learned decision policy.
\end{itemize}
\vspace{-4pt}
Low disagreement indicates that selective shard retraining closely approximates ideal full retraining, thereby validating unlearning correctness.

\subsubsection{Privacy Audit via MIA Proxy on Q-margins}
\label{subsubsec:mia_proxy}

A membership inference attack (MIA) tests whether a sample was used in training. Models often exhibit higher confidence in training samples than in unseen samples. Therefore, successful unlearning requires forgotten samples to become less distinguishable from never-observed data.

In this study, privacy leakage was assessed using a lightweight MIA proxy based on ensemble Q-value margins:
\begin{equation}
Q_{\mathrm{margin}}^{\mathrm{ens}}(s)
=
\frac{1}{M}
\sum_{m=1}^{M}
\left(
Q_{\theta_m}(s,1)-Q_{\theta_m}(s,0)
\right)
\label{eq:mia_qmargin}
\end{equation}
where \(M\) is the number of shards and \(Q_{\theta_m}\) is the Q-function of the shard \(m\). This scalar score reflects the confidence in the prediction. The distinguishability of the membership was quantified using ROC-AUC before and after deletion. The reduction in privacy leakage is measured as
\begin{equation}
\Delta AUC_{\mathrm{MIA}}
=
AUC_{\mathrm{MIA}}^{\mathrm{before}}-AUC_{\mathrm{MIA}}^{\mathrm{after}}
\label{eq:delta_auc_mia}
\end{equation}

A post-deletion AUC close to 0.5 indicates lower membership leakage and better privacy alignment.

\subsubsection{Computational Efficiency}
\label{subsubsec:computational_efficiency}

The computational efficiency of the DDQN--SISA framework was evaluated beyond detection accuracy and privacy alignment. As SISA-based unlearning aims to reduce the cost of retraining under deletion requests, system-level performance metrics were measured for each configuration. For every shard setting \(M\) and deletion scenario, we record:
\begin{itemize}
    \item \(T_{\mathrm{SISA}}\): full SISA training time across all shard-level DDQN models,
    \item \(T_{\mathrm{unlearning}}\): retraining time only for affected shards,
    \item \(T_{\mathrm{oracle}}\): oracle retraining time in the retained dataset \(\mathcal{D}_r\), and
    \item peak memory usage during training and retraining.
\end{itemize}

To assess computational efficiency, the communication overhead during shard-level retraining was estimated as
\begin{equation}
T_{\mathrm{comm}}
=
|\mathcal{M}_{\mathrm{aff}}|
\left(
\tau + 2\frac{B}{BW}
\right)
\label{eq:comm_overhead}
\end{equation}
where \(|\mathcal{M}_{\mathrm{aff}}|\) denotes the number of affected shards, \(\tau\) is the network latency per-shard, \(B\) is the size of the model in bytes and \(BW\) is the available bandwidth. The term \(2B/BW\) accounts for the transfer of bidirectional models (upload and download) during synchronization. This formulation shows that the cost of unlearning depends on the number of affected shards rather than the full dataset, supporting efficient repeated deletions.

To summarize the overall workflow and evaluation protocol described above, Algorithm~\ref{alg:ddqn_sisa} presents the complete end-to-end DDQN--SISA unlearning framework.

\begin{algorithm}[!htbp]
\caption{End-to-End DDQN--SISA Unlearning Framework}
\label{alg:ddqn_sisa}
\begin{algorithmic}[1]
\Require Dataset \(\mathcal{D}=\{(x_i,y_i)\}_{i=1..N}\), \(x_i\in\mathbb{R}^{103}\), \(y_i\in\{0,1\}\), \(N=2000\); stratified \(K\)-fold CV (\(K=5\)); RL agents \(\{DQN,DDQN\}\) with reward \(R2\) and \(\gamma\); shards \(M\in\{5,10,20\}\); rounds \(r\in\{1,5,10\}\); forget fractions \(f\in\{0.01,0.05,0.10\}\); scenarios \(SCEN\in\{\texttt{single\_shard},\texttt{multi\_shard}\}\); step budgets \(T_{\mathrm{base}},T_{\mathrm{sisa}},T_{\mathrm{unlearn}},T_{\mathrm{oracle}}\)
\Ensure Winner DDQN + full unlearning logs (Utility, Oracle disagreement, MIA proxy, Runtime, Comm overhead)

\Statex \textbf{Part A: Baseline RL Detector}
\State Fix global random seeds and enforce deterministic execution
\For{\(algo \in \{DQN,DDQN\}\)}
    \For{each fold \(k=1..5\)}
        \State Perform stratified split \(\mathcal{D} \rightarrow (\mathcal{D}_{tr}, \mathcal{D}_{te})\)
        \State Standardize using scaler fit on \(\mathcal{D}_{tr}\)
        \State Train \(algo\) on \(\mathcal{D}_{tr}\) for \(T_{\mathrm{base}}\) using reward \(R2\)
        \State Evaluate on \(\mathcal{D}_{te}\)
        \State Compute \(F1_k\) and Q-margin \(m(s)=Q(s,1)-Q(s,0)\) \(\rightarrow\) ROC/AUC
    \EndFor
    \State Compute mean \(F1\) over folds
\EndFor
\State Select winner \(A^*\) = model with the highest mean \(F1\) score (DDQN)

\Statex \textbf{Part B: DDQN--SISA Unlearning Suite}
\For{each shard count \(M \in \{5,10,20\}\)}
    \For{each fold \(k=1..5\)}
        \State Split and scale data as above
        \State \# SISA Training
        \State Partition \(\mathcal{D}_{tr}\) into \(M\) balanced shards \(\{\mathcal{D}_1,\ldots,\mathcal{D}_M\}\)
        \State Train one \(DDQN_m\) per shard for \(T_{\mathrm{sisa}}\)
        \State Aggregate predictions via majority vote over shard outputs
        \State Ensemble Q-margin \(m_{\mathrm{ens}}(s)=\mathrm{mean}_m(Q_m(s,1)-Q_m(s,0))\)
        \State Compute \(F1_{\mathrm{before}}\) on \(\mathcal{D}_{te}\)
        \State Store \(MIA\_AUC_{\mathrm{before}}\) using train vs.\ test margins

        \State \# Sequential Deletion
        \For{scenario \(sc \in \{\texttt{single\_shard}, \texttt{multi\_shard}\}\)}
            \State Define candidate pool \(\mathcal{C}\)
            \For{\(f \in \{0.01,0.05,0.10\}\)}
                \For{\(r \in \{1,5,10\}\)}
                    \State Select deletion set \(\mathcal{D}_f \subset \mathcal{C}\)
                    \State Retained set: \(\mathcal{D}_r = \mathcal{D}_{tr} \setminus \mathcal{D}_f\)
                    \State Affected shards: \(\mathcal{M}_{aff} = \{m : \mathcal{D}_m \cap \mathcal{D}_f \neq \emptyset\}\)

                    \State \# Unlearning = retrain affected shards only
                    \For{each \(m \in \mathcal{M}_{aff}\)}
                        \State Retrain \(DDQN_m\) on \(\mathcal{D}_m \setminus \mathcal{D}_f\) for \(T_{\mathrm{unlearn}}\) steps
                    \EndFor

                    \State \# Utility Preservation
                    \State Evaluate \(F1_{\mathrm{after}}\) on \(\mathcal{D}_{te}\)
                    \State \(\Delta F1 = F1_{\mathrm{after}} - F1_{\mathrm{before}}\), \(U_{\mathrm{drop}} = \max(0, F1_{\mathrm{before}} - F1_{\mathrm{after}})\)

                    \State \# Oracle Verification
                    \State Train oracle DDQN on \(\mathcal{D}_r\) for \(T_{\mathrm{oracle}}\)
                    \State Compute disagreement rates:
                    \State \hspace{1.6em}\(Dis_{\mathrm{forget}}\) (on \(\mathcal{D}_f\)), \(Dis_{\mathrm{retain}}\) (on \(\mathcal{D}_r\))

                    \State \# Membership Inference Proxy
                    \State Compute \(MIA\_AUC_{\mathrm{after}}\) using retained vs.\ test margins
                    \State \(\Delta AUC_{\mathrm{MIA}} = AUC_{\mathrm{before}} - AUC_{\mathrm{after}}\)

                    \State \# Computational Efficiency
                    \State Measure: \(T_{\mathrm{SISA}}\) (all shard training), \(T_{\mathrm{unlearn}}\), \(T_{\mathrm{oracle}}\), peak RAM
                    \State Estimate communication overhead:
                    \State \hspace{1.6em}\(T_{\mathrm{comm}} = |\mathcal{M}_{aff}| \cdot (\tau + 2\cdot B/BW)\), \(B\) = model size in bytes
                    \State Log all metrics \(\{M, sc, f, r, utility, oracle, MIA, runtime\}\)
                \EndFor
            \EndFor
        \EndFor
    \EndFor
\EndFor
\end{algorithmic}
\end{algorithm}

\section{Experimental Set Up}
All experiments were conducted in a CPU-only Google Colab Pro environment to ensure practical deployment under hardware constraints. The implementation used Python 3.10 with PyTorch 2.1 for value-based reinforcement learning and Scikit-learn 1.3 for stratified cross-validation and feature standardization. The results were reported using five-fold stratified cross-validation. The architectural design and training hyperparameters are summarized in Tables~\ref{tab:qnet_arch}--\ref{tab:eval_config}.

\begin{center}
\captionsetup{type=table}
\captionof{table}{Q-network architecture for DQN and DDQN.}
\label{tab:qnet_arch}
\footnotesize
\setlength{\tabcolsep}{5pt}
\begin{tabular}{@{}p{0.25\columnwidth} p{0.67\columnwidth}@{}}
\hline
\textbf{Layer} & \textbf{Configuration} \\
\hline
Input & 103 neurons (behavioral feature vector) \\
Hidden Layer 1 & 128 neurons, ReLU activation \\
Hidden Layer 2 & 128 neurons, ReLU activation \\
Output & 2 neurons representing Q-values for class 0 (benign) and class 1 (ransomware) \\
\hline
\end{tabular}
\end{center}

\begin{center}
\captionsetup{type=table}
\captionof{table}{Core training and SISA configuration.}
\label{tab:core_config}
\footnotesize
\setlength{\tabcolsep}{5pt}
\begin{tabular}{@{}p{0.34\columnwidth} p{0.60\columnwidth}@{}}
\hline
\textbf{Component} & \textbf{Configuration} \\
\hline
Optimizer & Adam optimizer with learning rate $\alpha = 0.001$ \\
Batch size & 64 samples per batch \\
Discount factor & $\gamma = 0.1$ \\
Replay buffer size & 50{,}000 experience tuples \\
Target network update & Every 500 training steps \\
Exploration strategy & $\epsilon$-greedy exploration from 1.0 to 0.05 with linear decay over 5{,}000 steps \\
Loss function & Smooth L1 (Huber) loss \\
Reward function ($R_2$) & +1 (correct), $-2$ (false negative), $-0.5$ (false positive) \\
Shard configurations & $M \in \{5,10,20\}$ \\
Deletion fractions & $\{1\%,5\%,10\%\}$ \\
Deletion rounds & $\{1,5,10\}$ \\
Scenarios & Single-shard and multi-shard deletion \\
Aggregation & Majority voting \\
\hline
\end{tabular}
\end{center}

\begin{center}
\captionsetup{type=table}
\captionof{table}{Unlearning budgets and evaluation metrics.}
\label{tab:eval_config}
\footnotesize
\setlength{\tabcolsep}{5pt}
\begin{tabular}{@{}p{0.36\columnwidth} p{0.58\columnwidth}@{}}
\hline
\textbf{Component} & \textbf{Configuration} \\
\hline
SISA training budget & 5{,}000 steps per shard \\
Unlearning retraining & 2{,}000 steps per affected shard \\
Oracle retraining & 5{,}000 steps on retained set $\mathcal{D}_r$ \\
Minimum retained shard size & 10 samples \\
Utility metrics & Accuracy, Precision, Recall, and F1-score \\
Utility change & $\Delta F1$ and $U_{\mathrm{drop}}$ \\
Oracle correctness metric & Disagreement on forget and retain sets \\
Privacy (MIA proxy) metric & ROC--AUC using ensemble Q-margin before and after deletion \\
Efficiency metrics & Training time ($T_{\mathrm{SISA}}, T_{\mathrm{unlearn}}, T_{\mathrm{oracle}}$) \\
Resource metric & Peak RAM usage (MB) \\
\hline
\end{tabular}
\end{center}

\section{Experimental Results}
\label{sec:experimental_results}
This section evaluates the DDQN-guided SISA unlearning framework across six dimensions: (i) baseline RL detection performance as a policy-learning benchmark, (ii) utility preservation under sequential deletion, (iii) oracle-verified forgetting correctness in terms of Q-policy alignment, (iv) privacy behavior under membership-inference analysis using Q-margin confidence signals, (v) computational efficiency under varying shard granularities and deletion severity, and (vi) contextual comparison with representative ransomware detection studies. All results were averaged over five-fold stratified cross-validation.

\subsection{Baseline Detection Performance}

To evaluate the effectiveness of the value-based learner, we first report the baseline detection performance using five-fold stratified cross-validation. Table~\ref{tab:baseline_detection} summarizes the baseline detection results prior to the unlearning operations. Both DQN and DDQN achieved F1-scores above 0.992 with Q-margin AUC values close to 0.999, indicating near-perfect discrimination between ransomware and benign behavior. DDQN achieved slightly higher mean performance and lower variance, suggesting improved training stability. Consequently, DDQN was selected as the base learner for the subsequent SISA unlearning experiments. The ROC curve in Fig.~\ref{fig:roc_ddqn} illustrates strong class separability, while inference latency remained negligible for both models; DDQN incurred only a modest increase in training and inference time.

\begin{center}
\captionsetup{type=table}
\captionof{table}{Baseline ransomware detection performance (non-SISA, DQN vs.\ DDQN).}
\label{tab:baseline_detection}
\footnotesize
\setlength{\tabcolsep}{3pt}
\resizebox{\columnwidth}{!}{%
\begin{tabular}{@{}lccccc@{}}
\hline
\textbf{Model} & \textbf{ID F1 (Mean \(\pm\) Std)} & \textbf{AUC (Q-score)} & \textbf{FN Rate} & \textbf{Train (s)} & \textbf{Infer (s)} \\
\hline
DQN  & \(0.9920 \pm 0.0045\) & 0.99874 & 0.008 & 30.20 & 0.0408 \\
DDQN & \(0.9925 \pm 0.0025\) & 0.99833 & 0.008 & 33.72 & 0.0440 \\
\hline
\end{tabular}%
}
\end{center}

\begin{center}
\captionsetup{type=figure}
\includegraphics[width=\columnwidth]{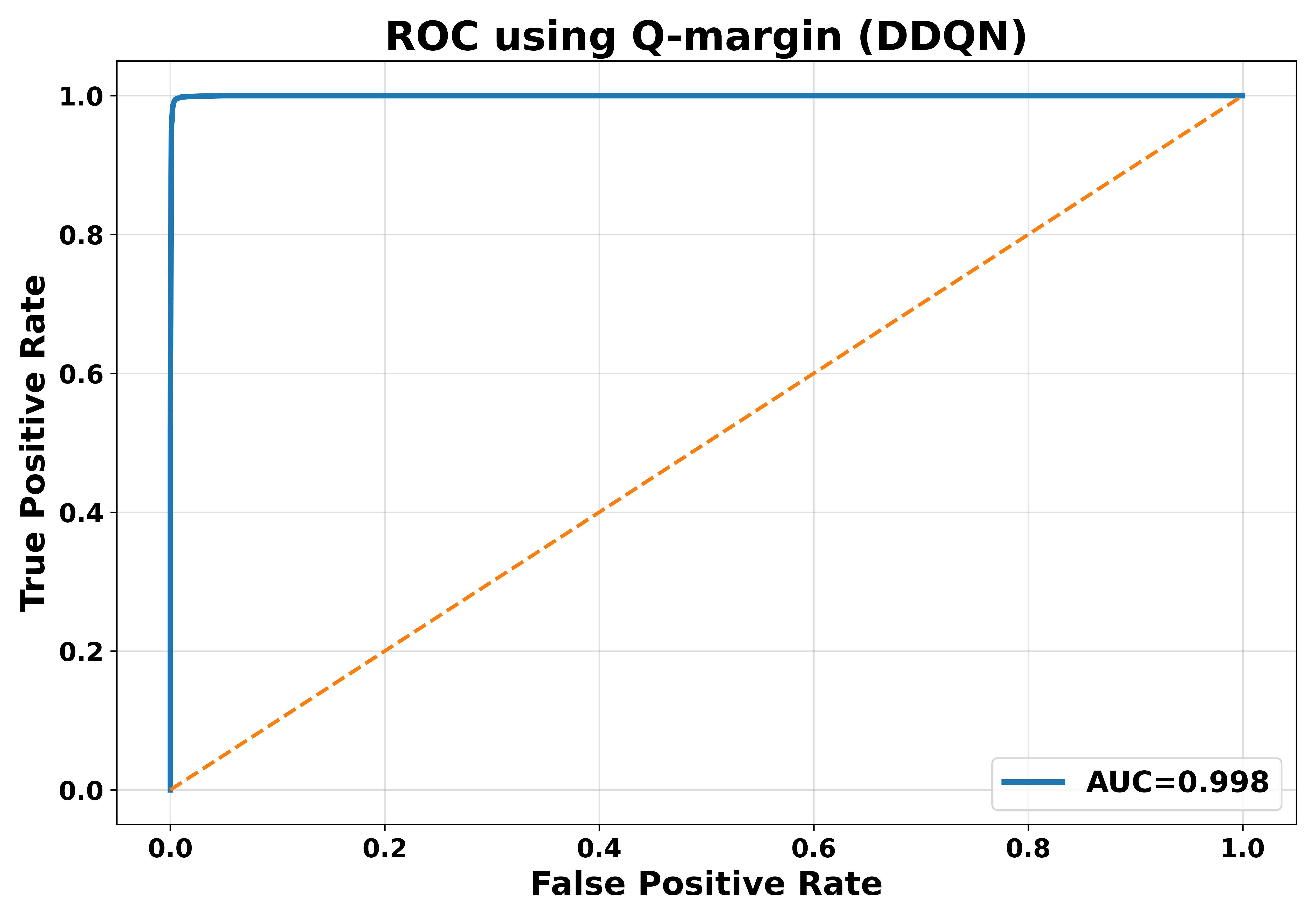}
\captionof{figure}{ROC curve using Q-margin (DDQN).}
\label{fig:roc_ddqn}
\end{center}

\subsection{Utility Preservation Under Sequential Unlearning}

Tables~\ref{tab:single_utility}--\ref{tab:multi_utility} and Fig.~\ref{fig:utility_heatmap} evaluate utility preservation under sequential deletion by measuring the performance change (\(|\Delta F1|\)) across shard granularities (\(M=5,10,20\)), deletion rounds (1, 5, 10), and forget fractions (1\%, 5\%, 10\%) for single- and multi-shard retraining. As illustrated in Fig.~\ref{fig:utility_heatmap}, the utility-drop heatmap reveals markedly different stability patterns between the single- and multi-shard SISA configurations. In single-shard settings, utility degradation remains minimal across all forget fractions and deletion rounds, with \(\Delta F1\) below 0.003 and only modest deviations under the most aggressive deletion conditions. This indicates strong robustness to sequential sample removal and limited sensitivity to increasing deletion rounds or forget fractions.

In contrast, multi-shard retraining exhibits substantially greater sensitivity to deletion intensity and shard size. At low deletion levels, particularly at a 1\% forget fraction, the performance remains close to that of single-shard retraining. However, as the forget fraction and deletion rounds increase, the degradation becomes increasingly pronounced, especially for \(M=10\) and \(M=20\). Under extreme stress conditions (10\% forget fraction and 10 rounds), the framework shows severe instability, with \(\Delta F1=-0.78207\) for \(M=10\) and \(\Delta F1=-0.95745\) for \(M=20\). This collapse is a predictable boundary condition rather than an implementation error. With \(M=20\) shards of approximately 100 samples each, a 10\% forget fraction applied over 10 cumulative rounds removes approximately 160 samples in total. When these deletions concentrate within already-small shards, the retained shard size can fall below the minimum viable threshold (10 samples in this configuration), at which point the affected shard model is excluded from the ensemble entirely. As progressively more shards are nullified, the ensemble loses representational diversity, and majority voting degrades toward random or biased outputs, producing the near-zero F1 scores observed. This behavior is consistent with the theoretical guarantee of SISA: shard-level isolation prevents deletion effects from propagating across shards, but it also means that a shard starved of data cannot self-correct from ensemble contributions. Consequently, \(M=20\) with aggressive deletion schedules (\(\geq 5\%\times 10\) rounds) is not recommended for deployment; moderate shard counts of \(M=5\)--10 provide the optimal balance between unlearning efficiency and policy stability under realistic regulatory deletion volumes.

Overall, the results show that SISA preserves utility well in single-shard settings and under moderate multi-shard configurations, whereas aggressive repeated deletions in high-granularity multi-shard settings can exhaust individual shard data budgets, triggering ensemble degradation that renders \(M=20\) unsuitable for deployment under sustained deletion pressure.

\begin{table*}[!t]
\caption{Single-shard utility preservation (mean \(\pm\) std over 5 folds).}
\label{tab:single_utility}
\centering
\scriptsize
\setlength{\tabcolsep}{3pt}
\renewcommand{\arraystretch}{0.90}
\resizebox{\textwidth}{!}{%
\begin{tabular}{@{}cccccccc@{}}
\hline
\textbf{M} & \textbf{Scenario} & \textbf{Forget \%} & \textbf{Rounds} & \textbf{F1 Before (\(\pm\) std)} & \textbf{F1 After (\(\pm\) std)} & \textbf{\(\Delta F1\) (\(\pm\) std)} & \textbf{Utility Drop} \\
\hline
5  & Single & 1\%  & 1  & \(0.97968 \pm 0.00316\) & \(0.97968 \pm 0.00316\) & \(+0.00000 \pm 0.00000\) & 0.00000 \\
5  & Single & 1\%  & 5  & \(0.97968 \pm 0.00316\) & \(0.97871 \pm 0.00279\) & \(-0.00098 \pm 0.00218\) & 0.00098 \\
5  & Single & 1\%  & 10 & \(0.97968 \pm 0.00316\) & \(0.97921 \pm 0.00439\) & \(-0.00047 \pm 0.00204\) & 0.00047 \\
5  & Single & 5\%  & 1  & \(0.97968 \pm 0.00316\) & \(0.97869 \pm 0.00444\) & \(-0.00099 \pm 0.00220\) & 0.00099 \\
5  & Single & 5\%  & 5  & \(0.97968 \pm 0.00316\) & \(0.97869 \pm 0.00277\) & \(-0.00100 \pm 0.00219\) & 0.00100 \\
5  & Single & 5\%  & 10 & \(0.97968 \pm 0.00316\) & \(0.97724 \pm 0.00316\) & \(-0.00245 \pm 0.00004\) & 0.00245 \\
5  & Single & 10\% & 1  & \(0.97968 \pm 0.00316\) & \(0.97917 \pm 0.00328\) & \(-0.00051 \pm 0.00114\) & 0.00051 \\
5  & Single & 10\% & 5  & \(0.97968 \pm 0.00316\) & \(0.97772 \pm 0.00242\) & \(-0.00197 \pm 0.00110\) & 0.00197 \\
5  & Single & 10\% & 10 & \(0.97968 \pm 0.00316\) & \(0.97156 \pm 0.00798\) & \(-0.00812 \pm 0.00899\) & 0.00812 \\
10 & Single & 1\%  & 1  & \(0.96802 \pm 0.00765\) & \(0.96565 \pm 0.00883\) & \(-0.00237 \pm 0.00236\) & 0.00237 \\
10 & Single & 1\%  & 5  & \(0.96802 \pm 0.00765\) & \(0.96563 \pm 0.00618\) & \(-0.00239 \pm 0.00338\) & 0.00239 \\
10 & Single & 1\%  & 10 & \(0.96802 \pm 0.00765\) & \(0.96616 \pm 0.00742\) & \(-0.00186 \pm 0.00200\) & 0.00186 \\
10 & Single & 5\%  & 1  & \(0.96802 \pm 0.00765\) & \(0.96707 \pm 0.00811\) & \(-0.00095 \pm 0.00130\) & 0.00095 \\
10 & Single & 5\%  & 5  & \(0.96802 \pm 0.00765\) & \(0.96568 \pm 0.00654\) & \(-0.00234 \pm 0.00245\) & 0.00234 \\
10 & Single & 5\%  & 10 & \(0.96802 \pm 0.00765\) & \(0.96620 \pm 0.00655\) & \(-0.00182 \pm 0.00319\) & 0.00182 \\
10 & Single & 10\% & 1  & \(0.96802 \pm 0.00765\) & \(0.96564 \pm 0.00744\) & \(-0.00238 \pm 0.00239\) & 0.00238 \\
10 & Single & 10\% & 5  & \(0.96802 \pm 0.00765\) & \(0.96524 \pm 0.00611\) & \(-0.00278 \pm 0.00322\) & 0.00278 \\
10 & Single & 10\% & 10 & \(0.96802 \pm 0.00765\) & \(0.97136 \pm 0.00659\) & \(+0.00334 \pm 0.00127\) & 0.00000 \\
20 & Single & 1\%  & 1  & \(0.95257 \pm 0.00207\) & \(0.95304 \pm 0.00172\) & \(+0.00047 \pm 0.00106\) & 0.00000 \\
20 & Single & 1\%  & 5  & \(0.95257 \pm 0.00207\) & \(0.95261 \pm 0.00211\) & \(+0.00004 \pm 0.00171\) & 0.00000 \\
20 & Single & 1\%  & 10 & \(0.95257 \pm 0.00207\) & \(0.95309 \pm 0.00299\) & \(+0.00052 \pm 0.00200\) & 0.00000 \\
20 & Single & 5\%  & 1  & \(0.95257 \pm 0.00207\) & \(0.95350 \pm 0.00345\) & \(+0.00093 \pm 0.00208\) & 0.00000 \\
20 & Single & 5\%  & 5  & \(0.95257 \pm 0.00207\) & \(0.95261 \pm 0.00332\) & \(+0.00004 \pm 0.00165\) & 0.00000 \\
20 & Single & 5\%  & 10 & \(0.95257 \pm 0.00207\) & \(0.95262 \pm 0.00202\) & \(+0.00005 \pm 0.00011\) & 0.00000 \\
20 & Single & 10\% & 1  & \(0.95257 \pm 0.00207\) & \(0.95258 \pm 0.00297\) & \(+0.00001 \pm 0.00165\) & 0.00000 \\
20 & Single & 10\% & 5  & \(0.95257 \pm 0.00207\) & \(0.95113 \pm 0.00293\) & \(-0.00144 \pm 0.00132\) & 0.00144 \\
20 & Single & 10\% & 10 & \(0.95257 \pm 0.00207\) & \(0.95468 \pm 0.00438\) & \(+0.00211 \pm 0.00248\) & 0.00000 \\
\hline
\end{tabular}%
}
\end{table*}

\begin{table*}[!t]
\caption{Utility preservation after multi-shard deletion (mean \(\pm\) std over 5 folds).}
\label{tab:multi_utility}
\centering
\scriptsize
\setlength{\tabcolsep}{3pt}
\renewcommand{\arraystretch}{0.90}
\resizebox{\textwidth}{!}{%
\begin{tabular}{@{}cccccccc@{}}
\hline
\textbf{M} & \textbf{Scenario} & \textbf{Forget \%} & \textbf{Rounds} & \textbf{F1 Before (\(\pm\) std)} & \textbf{F1 After (\(\pm\) std)} & \textbf{\(\Delta F1\) (\(\pm\) std)} & \textbf{Utility Drop} \\
\hline
5  & Multi & 1\%  & 1  & \(0.97816 \pm 0.00290\) & \(0.97714 \pm 0.00301\) & \(-0.00102 \pm 0.00205\) & 0.00102 \\
5  & Multi & 1\%  & 5  & \(0.97816 \pm 0.00290\) & \(0.97464 \pm 0.00487\) & \(-0.00352 \pm 0.00412\) & 0.00352 \\
5  & Multi & 1\%  & 10 & \(0.97816 \pm 0.00290\) & \(0.97577 \pm 0.00398\) & \(-0.00239 \pm 0.00325\) & 0.00239 \\
5  & Multi & 5\%  & 1  & \(0.97816 \pm 0.00290\) & \(0.97717 \pm 0.00302\) & \(-0.00099 \pm 0.00220\) & 0.00099 \\
5  & Multi & 5\%  & 5  & \(0.97816 \pm 0.00290\) & \(0.97619 \pm 0.00486\) & \(-0.00197 \pm 0.00486\) & 0.00197 \\
5  & Multi & 5\%  & 10 & \(0.97816 \pm 0.00290\) & \(0.95981 \pm 0.00876\) & \(-0.01835 \pm 0.00954\) & 0.01835 \\
5  & Multi & 10\% & 1  & \(0.97816 \pm 0.00290\) & \(0.97523 \pm 0.00347\) & \(-0.00293 \pm 0.00341\) & 0.00293 \\
5  & Multi & 10\% & 5  & \(0.97816 \pm 0.00290\) & \(0.95945 \pm 0.00912\) & \(-0.01870 \pm 0.00981\) & 0.01870 \\
5  & Multi & 10\% & 10 & \(0.97816 \pm 0.00290\) & \(0.90412 \pm 0.01086\) & \(-0.07404 \pm 0.01086\) & 0.07404 \\
10 & Multi & 1\%  & 1  & \(0.96661 \pm 0.00281\) & \(0.96814 \pm 0.00257\) & \(+0.00153 \pm 0.00248\) & 0.00000 \\
10 & Multi & 1\%  & 5  & \(0.96661 \pm 0.00281\) & \(0.96527 \pm 0.00293\) & \(-0.00134 \pm 0.00309\) & 0.00134 \\
10 & Multi & 1\%  & 10 & \(0.96661 \pm 0.00281\) & \(0.96201 \pm 0.00455\) & \(-0.00460 \pm 0.00472\) & 0.00460 \\
10 & Multi & 5\%  & 1  & \(0.96661 \pm 0.00281\) & \(0.96574 \pm 0.00321\) & \(-0.00087 \pm 0.00361\) & 0.00087 \\
10 & Multi & 5\%  & 5  & \(0.96661 \pm 0.00281\) & \(0.95533 \pm 0.00693\) & \(-0.01128 \pm 0.00693\) & 0.01128 \\
10 & Multi & 5\%  & 10 & \(0.96661 \pm 0.00281\) & \(0.94268 \pm 0.01119\) & \(-0.02393 \pm 0.01152\) & 0.02393 \\
10 & Multi & 10\% & 1  & \(0.96661 \pm 0.00281\) & \(0.96244 \pm 0.00394\) & \(-0.00417 \pm 0.00410\) & 0.00417 \\
10 & Multi & 10\% & 5  & \(0.96661 \pm 0.00281\) & \(0.94207 \pm 0.01108\) & \(-0.02453 \pm 0.01133\) & 0.02453 \\
10 & Multi & 10\% & 10 & \(0.96661 \pm 0.00281\) & \(0.18454 \pm 0.41309\) & \(-0.78207 \pm 0.41309\) & 0.78207 \\
20 & Multi & 1\%  & 1  & \(0.95745 \pm 0.00234\) & \(0.95417 \pm 0.00314\) & \(-0.00328 \pm 0.00351\) & 0.00328 \\
20 & Multi & 1\%  & 5  & \(0.95745 \pm 0.00234\) & \(0.94952 \pm 0.00472\) & \(-0.00793 \pm 0.00509\) & 0.00793 \\
20 & Multi & 1\%  & 10 & \(0.95745 \pm 0.00234\) & \(0.94892 \pm 0.00502\) & \(-0.00853 \pm 0.00541\) & 0.00853 \\
20 & Multi & 5\%  & 1  & \(0.95745 \pm 0.00234\) & \(0.95055 \pm 0.00428\) & \(-0.00690 \pm 0.00428\) & 0.00690 \\
20 & Multi & 5\%  & 5  & \(0.95745 \pm 0.00234\) & \(0.94753 \pm 0.00462\) & \(-0.00992 \pm 0.00462\) & 0.00992 \\
20 & Multi & 5\%  & 10 & \(0.95745 \pm 0.00234\) & \(0.93287 \pm 0.00871\) & \(-0.02458 \pm 0.00871\) & 0.02458 \\
20 & Multi & 10\% & 1  & \(0.95745 \pm 0.00234\) & \(0.94804 \pm 0.00435\) & \(-0.00941 \pm 0.00435\) & 0.00941 \\
20 & Multi & 10\% & 5  & \(0.95745 \pm 0.00234\) & \(0.93268 \pm 0.00883\) & \(-0.02477 \pm 0.00883\) & 0.02477 \\
20 & Multi & 10\% & 10 & \(0.95745 \pm 0.00234\) & \(0.00000 \pm 0.00000\) & \(-0.95745 \pm 0.00000\) & 0.95745 \\
\hline
\end{tabular}%
}
\end{table*}

\begin{figure*}[!t]
    \centering
    \includegraphics[width=\textwidth]{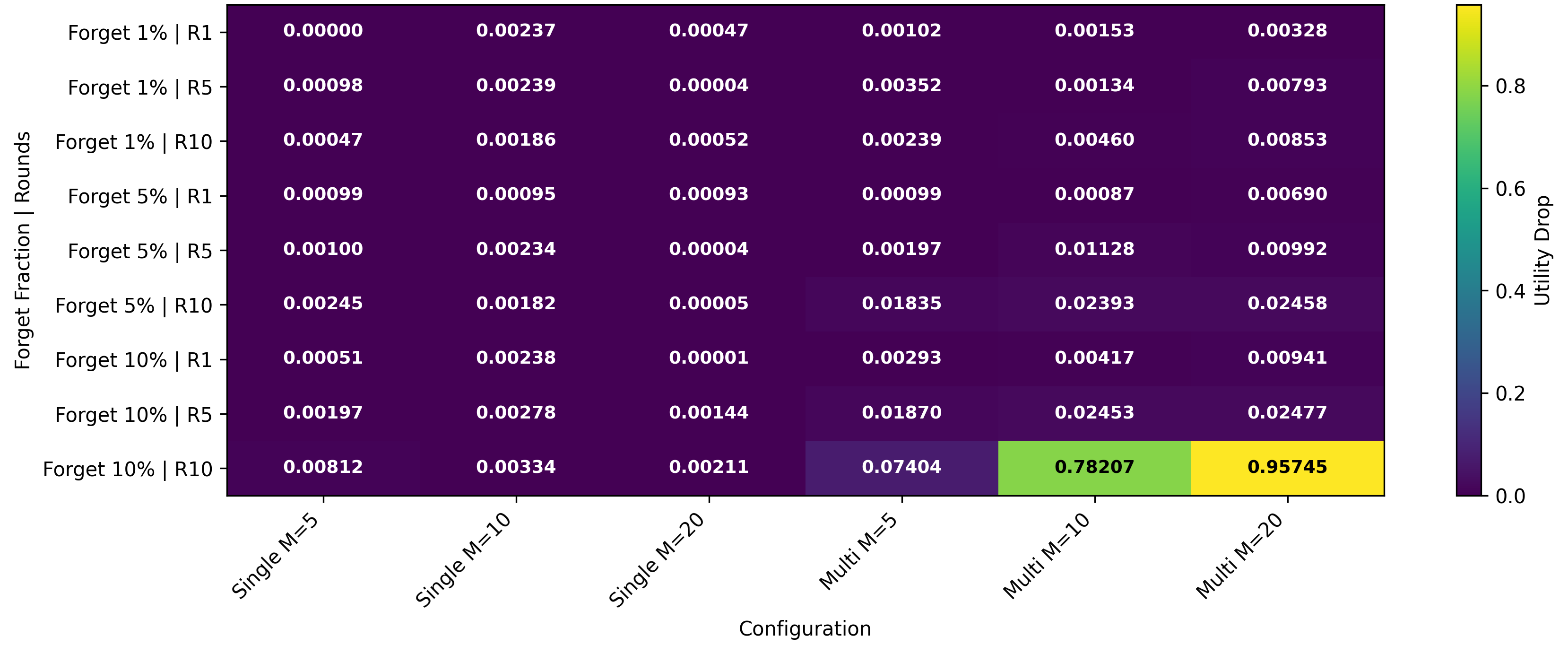}
    \caption{Heatmap of utility drop after machine unlearning across forget fractions (1\%, 5\%, and 10\%), deletion rounds (R1, R5, and R10), and SISA configurations (single- and multi-shard with \(M=\{5,10,20\}\)). Darker colors indicate lower performance degradation, whereas brighter colors indicate greater utility loss.}
    \label{fig:utility_heatmap}
\end{figure*}

\subsection{Oracle-Verified Forgetting}

Table~\ref{tab:oracle_forgetting} and Fig.~\ref{fig:oracle_disagreement} show the oracle disagreement rates (mean \(\pm\) standard deviation over five folds) for the forgotten and retained sets under single and multi-shard retraining. These rates quantify the deviation between the SISA-unlearned model and an oracle retrained from scratch on the retained data, thus indicating the correctness of unlearning. Single-shard retraining exhibited stable forgetting behavior across configurations. Forget-set disagreement increases with larger deletion fractions and repeated rounds but remains limited. For \(M=5\) and \(M=10\), the values remain below approximately 0.04 even under aggressive settings (10\% \(\times\) 10 rounds). Although \(M=20\) shows high variance at 1\% deletion \((0.200 \pm 0.447)\), this behavior does not persist under larger deletions. Retained-set disagreement also remains low (approximately 0.015--0.045) across single-shard settings, indicating that selective retraining preserves the decision behavior of non-deleted samples.

In contrast, multi-shard retraining becomes increasingly unstable when deletion pressure is high. For \(M=5\), the forget-set disagreement increases to 0.072 at 10\% \(\times\) 10 rounds. At higher shard counts, the instability becomes more pronounced: under 10\% \(\times\) 10 rounds, forget-set disagreement rises to 0.402 for \(M=10\) and 0.525 for \(M=20\), while retained-set disagreement simultaneously increases to 0.432 for \(M=10\) and 0.536 for \(M=20\). This indicates that repeated deletions across multiple shards can distort not only forgotten-sample behavior but also predictions on retained data. Overall, oracle verification reveals a clear stability trade-off: single-shard retraining supports controlled unlearning, whereas high-granularity multi-shard retraining is more vulnerable to error propagation under sustained deletion stress.

\begin{table*}[!t]
\caption{Oracle-verified forgetting for single-shard and multi-shard retraining (mean \(\pm\) std over 5 folds).}
\label{tab:oracle_forgetting}
\centering
\scriptsize
\setlength{\tabcolsep}{3pt}
\renewcommand{\arraystretch}{1.0}
\resizebox{\textwidth}{!}{%
\begin{tabular}{@{}ccccccc@{}}
\hline

\textbf{M} & \textbf{Forget \%} & \textbf{Rounds} & \textbf{Single-Shard Forget-set} & \textbf{Single-Shard Retained-set} & \textbf{Multi-Shard Forget-set} & \textbf{Multi-Shard Retained-set} \\
\hline
5  & 1\%  & 1  & \(0.000000 \pm 0.000000\) & \(0.015414 \pm 0.002285\) & \(0.012500 \pm 0.027951\) & \(0.017298 \pm 0.003238\) \\
5  & 1\%  & 5  & \(0.020000 \pm 0.027386\) & \(0.015443 \pm 0.002780\) & \(0.020000 \pm 0.014252\) & \(0.018158 \pm 0.002656\) \\
5  & 1\%  & 10 & \(0.025000 \pm 0.030619\) & \(0.014872 \pm 0.002097\) & \(0.023750 \pm 0.019466\) & \(0.018194 \pm 0.003345\) \\
5  & 5\%  & 1  & \(0.024265 \pm 0.033251\) & \(0.015662 \pm 0.002907\) & \(0.025000 \pm 0.017678\) & \(0.018421 \pm 0.002590\) \\
5  & 5\%  & 5  & \(0.021618 \pm 0.021435\) & \(0.015819 \pm 0.001906\) & \(0.016500 \pm 0.004541\) & \(0.018667 \pm 0.002007\) \\
5  & 5\%  & 10 & \(0.022868 \pm 0.012705\) & \(0.016589 \pm 0.003301\) & \(0.029000 \pm 0.006638\) & \(0.027500 \pm 0.004841\) \\
5  & 10\% & 1  & \(0.018004 \pm 0.027007\) & \(0.014165 \pm 0.002084\) & \(0.025000 \pm 0.015309\) & \(0.018056 \pm 0.001964\) \\
5  & 10\% & 5  & \(0.025482 \pm 0.012607\) & \(0.017125 \pm 0.002474\) & \(0.032250 \pm 0.009117\) & \(0.030000 \pm 0.006847\) \\
5  & 10\% & 10 & \(0.025240 \pm 0.012154\) & \(0.019282 \pm 0.003315\) & \(0.072258 \pm 0.018001\) & \(0.088000 \pm 0.017889\) \\

10 & 1\%  & 1  & \(0.000000 \pm 0.000000\) & \(0.026533 \pm 0.006720\) & \(0.075000 \pm 0.052291\) & \(0.030177 \pm 0.004052\) \\
10 & 1\%  & 5  & \(0.040000 \pm 0.054772\) & \(0.026541 \pm 0.005156\) & \(0.030000 \pm 0.006847\) & \(0.031053 \pm 0.004568\) \\
10 & 1\%  & 10 & \(0.030000 \pm 0.027386\) & \(0.026962 \pm 0.005897\) & \(0.047500 \pm 0.014389\) & \(0.032500 \pm 0.004484\) \\
10 & 5\%  & 1  & \(0.044444 \pm 0.060858\) & \(0.026895 \pm 0.005304\) & \(0.027500 \pm 0.010458\) & \(0.029737 \pm 0.003931\) \\
10 & 5\%  & 5  & \(0.027778 \pm 0.020031\) & \(0.027232 \pm 0.005707\) & \(0.043500 \pm 0.008404\) & \(0.037167 \pm 0.002674\) \\
10 & 5\%  & 10 & \(0.035278 \pm 0.017830\) & \(0.028277 \pm 0.005983\) & \(0.046250 \pm 0.005000\) & \(0.044750 \pm 0.006275\) \\
10 & 10\% & 1  & \(0.022876 \pm 0.031345\) & \(0.027030 \pm 0.005550\) & \(0.041250 \pm 0.018540\) & \(0.031806 \pm 0.004091\) \\
10 & 10\% & 5  & \(0.038471 \pm 0.009779\) & \(0.027961 \pm 0.005285\) & \(0.050500 \pm 0.008130\) & \(0.047750 \pm 0.007038\) \\
10 & 10\% & 10 & \(0.024929 \pm 0.010577\) & \(0.023661 \pm 0.006108\) & \(0.402452 \pm 0.195989\) & \(0.432000 \pm 0.219818\) \\

20 & 1\%  & 1  & \(0.200000 \pm 0.447214\) & \(0.043902 \pm 0.003891\) & \(0.037500 \pm 0.034233\) & \(0.043434 \pm 0.005996\) \\
20 & 1\%  & 5  & \(0.080000 \pm 0.109545\) & \(0.043511 \pm 0.002825\) & \(0.047500 \pm 0.027099\) & \(0.046316 \pm 0.004304\) \\
20 & 1\%  & 10 & \(0.060000 \pm 0.089443\) & \(0.044277 \pm 0.003404\) & \(0.042500 \pm 0.020444\) & \(0.046111 \pm 0.006165\) \\
20 & 5\%  & 1  & \(0.100000 \pm 0.136931\) & \(0.043740 \pm 0.003474\) & \(0.045000 \pm 0.022707\) & \(0.046974 \pm 0.005437\) \\
20 & 5\%  & 5  & \(0.050000 \pm 0.061237\) & \(0.044586 \pm 0.003958\) & \(0.053500 \pm 0.015472\) & \(0.051167 \pm 0.004023\) \\
20 & 5\%  & 10 & \(0.055000 \pm 0.062249\) & \(0.044419 \pm 0.003946\) & \(0.058250 \pm 0.007886\) & \(0.064500 \pm 0.006993\) \\
20 & 10\% & 1  & \(0.053571 \pm 0.073627\) & \(0.044969 \pm 0.003699\) & \(0.045000 \pm 0.028436\) & \(0.050000 \pm 0.002552\) \\
20 & 10\% & 5  & \(0.057857 \pm 0.048865\) & \(0.044588 \pm 0.003606\) & \(0.062500 \pm 0.004760\) & \(0.067250 \pm 0.004710\) \\
20 & 10\% & 10 & \(0.036071 \pm 0.014027\) & \(0.042219 \pm 0.002165\) & \(0.525032 \pm 0.014615\) & \(0.536000 \pm 0.071274\) \\
\hline
\end{tabular}%
}
\end{table*}

\begin{figure*}[!t]
    \centering
    \includegraphics[width=\textwidth]{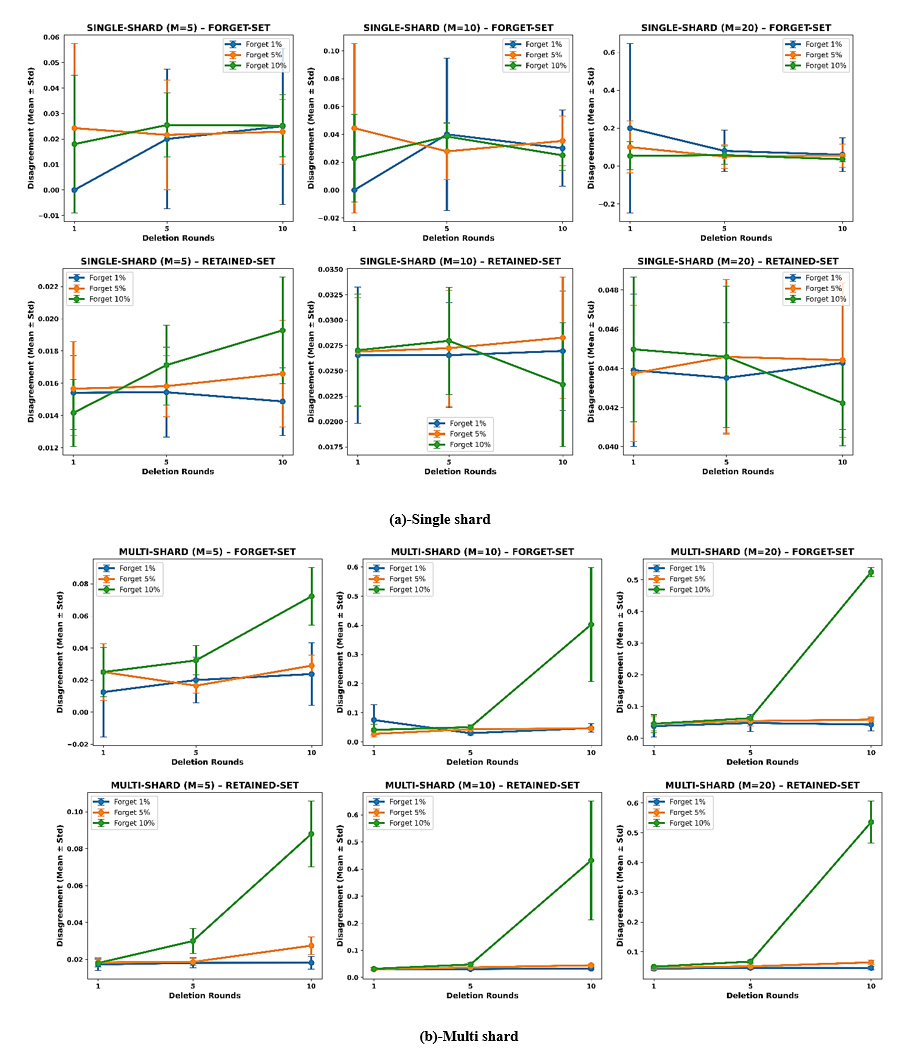}
    \caption{Oracle disagreement (mean \(\pm\) std over five folds) under single-shard (a) and multi-shard (b) retraining for shard counts \(M=\{5,10,20\}\), showing forget-set (top row) and retained-set (bottom row) disagreements across forget fractions (1\%, 5\%, 10\%) and deletion rounds (\(R=1,5,10\)).}
    \label{fig:oracle_disagreement}
\end{figure*}


\subsection{Membership-Inference Risk Analysis}

Table~\ref{tab:mia_analysis} and Fig.~\ref{fig:mia_delta} evaluate the privacy behavior between shard counts (\(M\in\{5,10,20\}\)), deletion fractions (1\%, 5\% and 10\%) and sequential deletion rounds (1, 5 and 10). Here, \(\Delta MIA\) is computed as the change in attack success after unlearning relative to before unlearning and is averaged over five folds. A zero value indicates no change in privacy, negative values indicate reduced attack success (privacy improvement), and positive values indicate increased distinguishability.

For single-shard unlearning, \(\Delta MIA\) remains close to zero in all settings, indicating that repeated deletions do not materially alter the privacy of the membership. Multi-shard unlearning shows a similar pattern under low and moderate deletion levels, with only small positive shifts under the most aggressive setting. Even in these cases, the deviations remain limited and do not exhibit the type of rapid escalation observed in the utility and oracle-disagreement analyses. Overall, the results indicate that the proposed SISA-based unlearning framework largely preserves membership privacy across practical configurations, with only minor degradation under extreme deletion stress.

\begin{table*}[!t]
\caption{Membership-inference attack success before and after unlearning (single-shard and multi-shard, mean \(\pm\) std over 5 folds).}
\label{tab:mia_analysis}
\centering
\scriptsize
\setlength{\tabcolsep}{3pt}
\renewcommand{\arraystretch}{0.92}
\resizebox{\textwidth}{!}{%
\begin{tabular}{@{}ccccccccc@{}}
\hline
\rule{0pt}{2.8ex}
\textbf{M} & \textbf{Forget \%} & \textbf{Rounds} &
\multicolumn{3}{c}{\textbf{Single-Shard}} &
\multicolumn{3}{c}{\textbf{Multi-Shard}} \\
\cline{4-6} \cline{7-9}
\rule{0pt}{2.6ex}
& & &
\textbf{MIA Before} & \textbf{MIA After} & \textbf{\(\Delta MIA\)} &
\textbf{MIA Before} & \textbf{MIA After} & \textbf{\(\Delta MIA\)} \\
\hline
5  & 1\%  & 1  & \(0.500 \pm 0.005\) & \(0.501 \pm 0.005\) & \(+0.001 \pm 0.002\) & \(0.502 \pm 0.006\) & \(0.497 \pm 0.004\) & \(-0.004 \pm 0.002\) \\
5  & 1\%  & 5  & \(0.500 \pm 0.005\) & \(0.499 \pm 0.004\) & \(-0.001 \pm 0.003\) & \(0.502 \pm 0.006\) & \(0.499 \pm 0.005\) & \(-0.002 \pm 0.003\) \\
5  & 1\%  & 10 & \(0.500 \pm 0.005\) & \(0.500 \pm 0.006\) & \(+0.000 \pm 0.002\) & \(0.502 \pm 0.006\) & \(0.500 \pm 0.004\) & \(-0.001 \pm 0.004\) \\
5  & 5\%  & 1  & \(0.500 \pm 0.005\) & \(0.501 \pm 0.005\) & \(+0.001 \pm 0.002\) & \(0.502 \pm 0.006\) & \(0.499 \pm 0.006\) & \(-0.002 \pm 0.004\) \\
5  & 5\%  & 5  & \(0.500 \pm 0.005\) & \(0.501 \pm 0.005\) & \(+0.001 \pm 0.004\) & \(0.502 \pm 0.006\) & \(0.502 \pm 0.006\) & \(+0.000 \pm 0.007\) \\
5  & 5\%  & 10 & \(0.500 \pm 0.005\) & \(0.500 \pm 0.006\) & \(-0.000 \pm 0.004\) & \(0.502 \pm 0.006\) & \(0.505 \pm 0.010\) & \(+0.003 \pm 0.013\) \\
5  & 10\% & 1  & \(0.500 \pm 0.005\) & \(0.499 \pm 0.006\) & \(-0.001 \pm 0.002\) & \(0.502 \pm 0.006\) & \(0.499 \pm 0.007\) & \(-0.002 \pm 0.006\) \\
5  & 10\% & 5  & \(0.500 \pm 0.005\) & \(0.500 \pm 0.006\) & \(+0.000 \pm 0.004\) & \(0.502 \pm 0.006\) & \(0.504 \pm 0.013\) & \(+0.003 \pm 0.014\) \\
5  & 10\% & 10 & \(0.500 \pm 0.005\) & \(0.499 \pm 0.005\) & \(-0.000 \pm 0.004\) & \(0.502 \pm 0.006\) & \(0.522 \pm 0.037\) & \(+0.021 \pm 0.041\) \\

10 & 1\%  & 1  & \(0.501 \pm 0.006\) & \(0.499 \pm 0.005\) & \(-0.001 \pm 0.003\) & \(0.501 \pm 0.007\) & \(0.499 \pm 0.004\) & \(-0.002 \pm 0.004\) \\
10 & 1\%  & 5  & \(0.501 \pm 0.006\) & \(0.500 \pm 0.004\) & \(-0.001 \pm 0.004\) & \(0.501 \pm 0.007\) & \(0.499 \pm 0.006\) & \(-0.002 \pm 0.004\) \\
10 & 1\%  & 10 & \(0.501 \pm 0.006\) & \(0.500 \pm 0.005\) & \(-0.001 \pm 0.005\) & \(0.501 \pm 0.007\) & \(0.499 \pm 0.006\) & \(-0.002 \pm 0.005\) \\
10 & 5\%  & 1  & \(0.501 \pm 0.006\) & \(0.500 \pm 0.005\) & \(-0.001 \pm 0.003\) & \(0.501 \pm 0.007\) & \(0.499 \pm 0.004\) & \(-0.002 \pm 0.006\) \\
10 & 5\%  & 5  & \(0.501 \pm 0.006\) & \(0.499 \pm 0.004\) & \(-0.002 \pm 0.004\) & \(0.501 \pm 0.007\) & \(0.501 \pm 0.004\) & \(-0.000 \pm 0.006\) \\
10 & 5\%  & 10 & \(0.501 \pm 0.006\) & \(0.500 \pm 0.004\) & \(-0.001 \pm 0.003\) & \(0.501 \pm 0.007\) & \(0.501 \pm 0.006\) & \(+0.000 \pm 0.007\) \\
10 & 10\% & 1  & \(0.501 \pm 0.006\) & \(0.500 \pm 0.004\) & \(-0.001 \pm 0.002\) & \(0.501 \pm 0.007\) & \(0.500 \pm 0.004\) & \(-0.001 \pm 0.006\) \\
10 & 10\% & 5  & \(0.501 \pm 0.006\) & \(0.499 \pm 0.004\) & \(-0.002 \pm 0.005\) & \(0.501 \pm 0.007\) & \(0.498 \pm 0.005\) & \(-0.003 \pm 0.005\) \\
10 & 10\% & 10 & \(0.501 \pm 0.006\) & \(0.501 \pm 0.006\) & \(+0.000 \pm 0.004\) & \(0.501 \pm 0.007\) & \(0.515 \pm 0.036\) & \(+0.014 \pm 0.038\) \\

20 & 1\%  & 1  & \(0.500 \pm 0.005\) & \(0.500 \pm 0.004\) & \(-0.000 \pm 0.004\) & \(0.501 \pm 0.004\) & \(0.500 \pm 0.003\) & \(-0.001 \pm 0.004\) \\
20 & 1\%  & 5  & \(0.500 \pm 0.005\) & \(0.499 \pm 0.003\) & \(-0.001 \pm 0.004\) & \(0.501 \pm 0.004\) & \(0.500 \pm 0.003\) & \(-0.001 \pm 0.003\) \\
20 & 1\%  & 10 & \(0.500 \pm 0.005\) & \(0.500 \pm 0.004\) & \(+0.000 \pm 0.004\) & \(0.501 \pm 0.004\) & \(0.499 \pm 0.004\) & \(-0.002 \pm 0.004\) \\
20 & 5\%  & 1  & \(0.500 \pm 0.005\) & \(0.499 \pm 0.003\) & \(-0.001 \pm 0.003\) & \(0.501 \pm 0.004\) & \(0.500 \pm 0.003\) & \(-0.001 \pm 0.003\) \\
20 & 5\%  & 5  & \(0.500 \pm 0.005\) & \(0.500 \pm 0.004\) & \(-0.000 \pm 0.003\) & \(0.501 \pm 0.004\) & \(0.500 \pm 0.003\) & \(-0.001 \pm 0.003\) \\
20 & 5\%  & 10 & \(0.500 \pm 0.005\) & \(0.500 \pm 0.004\) & \(+0.000 \pm 0.003\) & \(0.501 \pm 0.004\) & \(0.499 \pm 0.003\) & \(-0.001 \pm 0.004\) \\
20 & 10\% & 1  & \(0.500 \pm 0.005\) & \(0.501 \pm 0.003\) & \(+0.001 \pm 0.002\) & \(0.501 \pm 0.004\) & \(0.500 \pm 0.003\) & \(-0.001 \pm 0.006\) \\
20 & 10\% & 5  & \(0.500 \pm 0.005\) & \(0.499 \pm 0.004\) & \(-0.001 \pm 0.003\) & \(0.501 \pm 0.004\) & \(0.500 \pm 0.003\) & \(-0.001 \pm 0.004\) \\
20 & 10\% & 10 & \(0.500 \pm 0.005\) & \(0.500 \pm 0.004\) & \(+0.000 \pm 0.002\) & \(0.501 \pm 0.004\) & \(0.500 \pm 0.000\) & \(-0.001 \pm 0.004\) \\
\hline
\end{tabular}%
}
\end{table*}


\begin{figure}[!htbp]
    \centering
    \includegraphics[width=0.55\columnwidth]{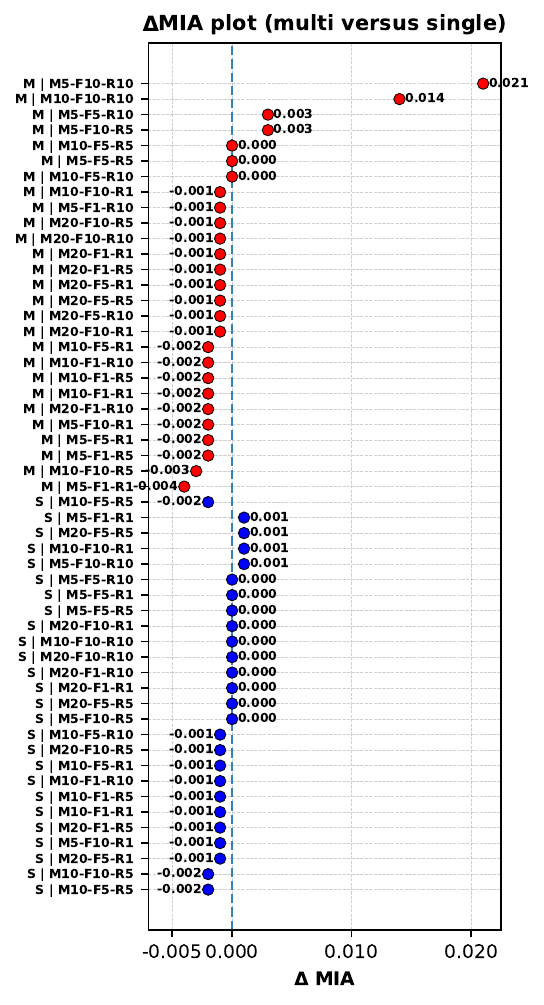}
\captionof{figure}{Membership-inference privacy change (\(\Delta MIA\)) across single- and multi-shard unlearning configurations under varying shard counts, forget fractions, and deletion rounds. Values near zero indicate negligible change in attack success after unlearning, whereas positive deviations under extreme deletion settings indicate modest increases in membership distinguishability.}
\label{fig:mia_delta}
\end{figure}

\subsection{Computational Efficiency Analysis}

Table~\ref{tab:efficiency_analysis} and Fig.~\ref{fig:efficiency_contour} summarize the computational behavior of the DDQN-based SISA framework using \(\Delta F1\), the full SISA training time, the unlearning time, and the communication overhead. The complete SISA training time scales proportionally with the shard count \(M\), from 83 \, s (\(M=5\)) to 326 \, s (\(M=20\)), reflecting the cost of training independent DDQN models \(M\). However, this cost is incurred only once during the initial construction of the SISA model.

By contrast, selective unlearning is substantially faster than full retraining. Single-shard retraining remains generally low and stable, typically at approximately 5.6--6.5\,s across most configurations, indicating strong computational locality and limited sensitivity to deletion intensity. Multi-shard retraining is more expensive because multiple shards may be affected simultaneously, with typical unlearning times of approximately 29--30\,s for \(M=5\), 23--24\,s for \(M=10\), and 20--21\,s for \(M=20\) under moderate deletions. Under aggressive deletions (10\% \(\times\) 10 rounds), the multi-shard retraining time decreases to approximately 10--18\,s, likely because fewer retained samples remain within the affected shards.

Utility trends reveal a corresponding trade-off. Single-shard retraining preserves performance with near-zero \(\Delta F1\) in most settings, whereas multi-shard configurations remain stable mainly under mild deletions and degrade sharply under aggressive repeated deletions, particularly at higher shard counts. Communication overhead scales with the number of affected shards, remaining at approximately 11\,ms for single-shard settings and approximately 53--55\,ms for multi-shard settings. In general, selective retraining of shards is substantially more efficient than full retraining, and moderate shard counts (\(M=5\) --10) provide the most favorable balance between efficiency and stability.

\begin{table*}[!t]
\caption{Efficiency analysis of DDQN-based SISA unlearning under varying shard granularity.}
\label{tab:efficiency_analysis}
\centering
\scriptsize
\setlength{\tabcolsep}{4pt}
\renewcommand{\arraystretch}{1.05}
\resizebox{\textwidth}{!}{%
\begin{tabular}{@{}ccccccccccc@{}}
\hline
\rule{0pt}{3.0ex}
\textbf{M} & \textbf{Forget \%} & \textbf{Rounds} &
\multicolumn{4}{c}{\textbf{Single-shard}} &
\multicolumn{4}{c}{\textbf{Multi-shard}} \\
\cline{4-7} \cline{8-11}
\rule{0pt}{2.8ex}
& & &
\shortstack{\textbf{\(\Delta F1\)}\\\textbf{(Mean \(\pm\) Std)}} &
\shortstack{\textbf{SISA Full}\\\textbf{Train (s)}} &
\shortstack{\textbf{Unlearn}\\\textbf{Time (s)}} &
\shortstack{\textbf{Comm}\\\textbf{Overhead (ms)}} &
\shortstack{\textbf{\(\Delta F1\)}\\\textbf{(Mean \(\pm\) Std)}} &
\shortstack{\textbf{SISA Full}\\\textbf{Train (s)}} &
\shortstack{\textbf{Unlearn}\\\textbf{Time (s)}} &
\shortstack{\textbf{Comm}\\\textbf{Overhead (ms)}} \\
\hline
5  & 1\%  & 1  & \(+0.0000 \pm 0.0000\) & 83.31  & 6.48 & 11.01 & \(-0.00102 \pm 0.00205\) & 80.78  & 29.17 & 52.84 \\
5  & 1\%  & 5  & \(-0.0010 \pm 0.0022\) & 83.31  & 6.38 & 11.01 & \(-0.00352 \pm 0.00412\) & 80.78  & 30.01 & 55.04 \\
5  & 1\%  & 10 & \(-0.0005 \pm 0.0020\) & 83.31  & 6.26 & 11.01 & \(-0.00239 \pm 0.00325\) & 80.78  & 30.48 & 55.04 \\
5  & 5\%  & 1  & \(-0.0010 \pm 0.0022\) & 83.31  & 6.25 & 11.01 & \(-0.00099 \pm 0.00220\) & 80.78  & 29.79 & 55.04 \\
5  & 5\%  & 5  & \(-0.0010 \pm 0.0022\) & 83.31  & 6.35 & 11.01 & \(-0.00197 \pm 0.00486\) & 80.78  & 29.59 & 55.04 \\
5  & 5\%  & 10 & \(-0.0024 \pm 0.0000\) & 83.31  & 6.47 & 11.01 & \(-0.01835 \pm 0.00954\) & 80.78  & 29.73 & 55.04 \\
5  & 10\% & 1  & \(-0.0005 \pm 0.0011\) & 83.31  & 6.49 & 11.01 & \(-0.00293 \pm 0.00341\) & 80.78  & 29.95 & 55.04 \\
5  & 10\% & 5  & \(-0.0020 \pm 0.0011\) & 83.31  & 6.41 & 11.01 & \(-0.01870 \pm 0.00981\) & 80.78  & 29.77 & 55.04 \\
5  & 10\% & 10 & \(-0.0081 \pm 0.0090\) & 83.31  & 6.45 & 11.01 & \(-0.07404 \pm 0.01086\) & 80.78  & 17.65 & 55.04 \\

10 & 1\%  & 1  & \(-0.0024 \pm 0.0024\) & 154.09 & 5.74 & 11.01 & \(+0.00153 \pm 0.00248\) & 154.79 & 23.52 & 55.04 \\
10 & 1\%  & 5  & \(-0.0024 \pm 0.0034\) & 154.09 & 5.80 & 11.01 & \(-0.00134 \pm 0.00309\) & 154.79 & 23.35 & 55.04 \\
10 & 1\%  & 10 & \(-0.0005 \pm 0.0024\) & 154.09 & 5.89 & 11.01 & \(-0.00460 \pm 0.00472\) & 154.79 & 23.40 & 55.04 \\
10 & 5\%  & 1  & \(-0.0010 \pm 0.0015\) & 154.09 & 5.87 & 11.01 & \(-0.00087 \pm 0.00361\) & 154.79 & 23.43 & 55.04 \\
10 & 5\%  & 5  & \(-0.0023 \pm 0.0024\) & 154.09 & 6.15 & 11.01 & \(-0.01128 \pm 0.00693\) & 154.79 & 23.24 & 55.04 \\
10 & 5\%  & 10 & \(+0.0038 \pm 0.0013\) & 154.09 & 6.12 & 11.01 & \(-0.02393 \pm 0.01152\) & 154.79 & 23.31 & 55.04 \\
10 & 10\% & 1  & \(+0.0017 \pm 0.0015\) & 154.09 & 6.01 & 11.01 & \(-0.00417 \pm 0.00410\) & 154.79 & 23.54 & 55.04 \\
10 & 10\% & 5  & \(-0.0033 \pm 0.0013\) & 154.09 & 6.10 & 11.01 & \(-0.02453 \pm 0.01133\) & 154.79 & 23.60 & 55.04 \\
10 & 10\% & 10 & \(+0.0033 \pm 0.0013\) & 154.09 & 6.25 & 11.01 & \(-0.78207 \pm 0.41309\) & 154.79 & 10.52 & 55.04 \\

20 & 1\%  & 1  & \(-0.0010 \pm 0.0020\) & 326.27 & 5.66 & 11.01 & \(-0.00328 \pm 0.00351\) & 329.79 & 20.60 & 55.04 \\
20 & 1\%  & 5  & \(+0.0004 \pm 0.0011\) & 326.27 & 5.85 & 11.01 & \(-0.00793 \pm 0.00509\) & 329.79 & 20.61 & 55.04 \\
20 & 1\%  & 10 & \(-0.0011 \pm 0.0016\) & 326.27 & 5.64 & 11.01 & \(-0.00853 \pm 0.00541\) & 329.79 & 20.75 & 55.04 \\
20 & 5\%  & 1  & \(+0.0009 \pm 0.0020\) & 326.27 & 5.70 & 11.01 & \(-0.00690 \pm 0.00428\) & 329.79 & 20.60 & 55.04 \\
20 & 5\%  & 5  & \(+0.0000 \pm 0.0016\) & 326.27 & 5.63 & 11.01 & \(-0.00992 \pm 0.00462\) & 329.79 & 20.59 & 55.04 \\
20 & 5\%  & 10 & \(+0.0039 \pm 0.0025\) & 326.27 & 5.73 & 11.01 & \(-0.02458 \pm 0.00871\) & 329.79 & 11.02 & 55.04 \\
20 & 10\% & 1  & \(+0.0004 \pm 0.0011\) & 326.27 & 5.80 & 11.01 & \(-0.00941 \pm 0.00435\) & 329.79 & 20.61 & 55.04 \\
20 & 10\% & 5  & \(-0.0021 \pm 0.0025\) & 326.27 & 5.73 & 11.01 & \(-0.02477 \pm 0.00883\) & 329.79 & 11.02 & 55.04 \\
20 & 10\% & 10 & \(+0.0021 \pm 0.0025\) & 326.27 & 5.66 & 11.01 & \(-0.95745 \pm 0.00000\) & 329.79 & 11.02 & 55.04 \\
\hline
\end{tabular}%
}
\end{table*}

\begin{figure*}[!t]
    \centering
    \includegraphics[width=\textwidth]{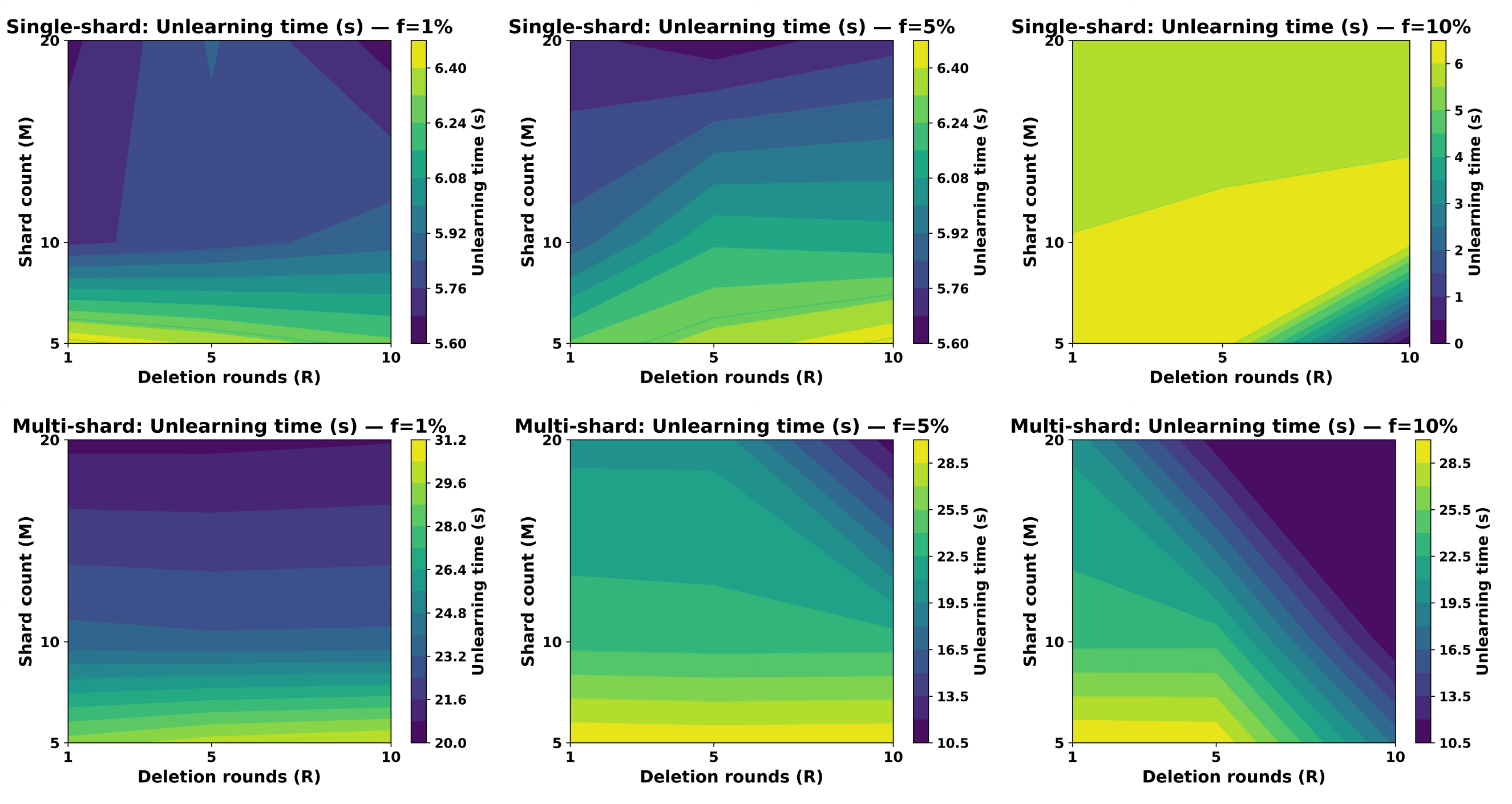}
    \caption{Contour visualization of unlearning time across shard granularities (\(M\)) and deletion rounds (\(R\)) for different forget fractions (\(f\)). The top and bottom rows show single- and multi-shard retraining. The results indicate that single-shard retraining maintains a stable cost, whereas multi-shard retraining is more sensitive to deletion intensity.}
    \label{fig:efficiency_contour}
\end{figure*}

\FloatBarrier

\subsection{Comparison with State-of-the-Art Ransomware Detection Performance}

Table~\ref{tab:sota_comparison} compares the proposed framework with representative ransomware detection studies. Because prior studies used different datasets and experimental protocols, the comparison is intended for contextual positioning rather than direct benchmarking.

\begin{table*}[!t]
\caption{Comparison with representative state-of-the-art ransomware detection studies.}
\label{tab:sota_comparison}
\centering
\scriptsize
\setlength{\tabcolsep}{5pt}
\renewcommand{\arraystretch}{1.08}
\resizebox{\textwidth}{!}{%
\begin{tabular}{@{}p{0.08\textwidth} p{0.47\textwidth} p{0.18\textwidth} p{0.19\textwidth}@{}}
\hline
\rule{0pt}{2.8ex}
\textbf{Ref.} & \textbf{Method / Features} & \textbf{Performance} & \textbf{Unlearning Support} \\
\hline
{[24]} & Memory-forensics features + PSO-based ensemble & \(\sim\)99\% & No \\
{[25]} & Dynamic behavioral features + CVAE-1D CNN & \(F1=94.62\%\) & No \\
{[23]} & ETW file I/O patterns + model-agnostic ML & \(Acc=99.69\%\) & No \\
{[26]} & API, DLL, and mutex traces + CNN & \(Acc=98.2\%\) & No \\
{[20]} & System-call groups + RLR/RF & \(Acc=97.48\%\) & No \\
{[29]} & Android permissions/network traffic + DRL & \(F1=91.20\%\) & No \\
{[28]} & PE-header features + deep RL & \(F1=97.7\%\) & No \\
Ours & Dynamic behavioral features + DDQN--SISA & \(F1=99.25\%\) & Yes: Unlearning + Oracle + MIA \\
\hline
\end{tabular}%
}
\end{table*}

As shown in Table~\ref{tab:sota_comparison}, recent ransomware detection systems have reported strong predictive performance across diverse feature modalities, including memory forensics, event tracing, API/DLL/mutex traces, behavior-based system calls, latent behavioral representations, and static PE-header features. However, these methods focus primarily on detection accuracy and assume fixed training data after deployment without addressing the requirement to remove previously learned samples under privacy or compliance constraints. RL-based studies similarly emphasize early detection or adaptive classification behavior rather than model updateability under deletion requests.

The proposed framework maintains a competitive baseline detection performance, with DDQN achieving \(F1>0.992\) and Q-margin AUC close to 0.999, while introducing capabilities not present in the compared ransomware detectors: multi-shard SISA retraining, oracle-verified forgetting, membership-inference privacy auditing, and scalability analysis under deletion stress. The experimental results further show that utility preservation remains strong under single-shard retraining and moderate multi-shard configurations (\(M=5\)--10), whereas larger shard counts (\(M=20\)) exhibit substantially greater instability under aggressive repeated deletions. Thus, rather than claiming uniform robustness across all settings, the results identify a practical operating regime in which privacy-compliant sample removal can be achieved with limited utility loss and moderate retraining costs.

While existing ransomware detection methods primarily emphasize classification performance, the proposed framework advances the field by integrating RL-based detection with SISA-based machine unlearning and audit-oriented verification. To the best of our knowledge, this is the first ransomware detection framework to jointly combine value-based RL, shard-local machine unlearning, oracle-based forgetting validation, and membership-inference privacy auditing within a single evaluation pipeline.

\section{Discussion}

This section interprets the experimental findings presented in Section~\ref{sec:experimental_results} and discusses their implications for DRL-based ransomware detection systems that require auditable and privacy-compliant machine unlearning.

This study investigated the integration of multi-shard SISA with DDQN-based ransomware detection for privacy-compliant model updating. The results show that the proposed framework can maintain strong baseline detection performance while enabling selective removal of training samples through localized shard-level retraining.

Prior ransomware detection approaches have largely relied on supervised ML and DL models without addressing data deletion or privacy requirements~\cite{ref3,ref4}. Unlike existing SISA-based unlearning frameworks developed primarily for supervised learning~\cite{ref9,ref10,ref12,ref13}, the proposed framework extends shard-based unlearning to a value-based reinforcement learning setting while incorporating oracle verification and membership-inference-based privacy auditing. This combination enables the assessment of both deletion correctness and post-unlearning privacy behavior.

The experimental results reveal a clear stability--efficiency trade-off across shard configurations. Single-shard retraining maintains high utility stability with only minimal degradation in most settings. Multi-shard configurations with moderate shard counts (\(M=5\)--10) remain effective under low-to-moderate deletion pressure and provide a favorable balance between retraining efficiency and predictive stability. However, a higher granularity of the shards (\(M=20\)) under repeated aggressive deletions introduces substantially greater utility loss and oracle disagreement. This behavior is likely due to fragmented data allocation, reduced per-shard sample diversity, and increased policy perturbations during localized retraining. Thus, shard granularity is not only a computational design choice but also a critical factor affecting unlearning reliability.

Oracle-based evaluation further shows that configurations with low disagreement preserve predictive behavior on retained data while removing the influence of forgotten samples. This supports the correctness of localized retraining as an approximation of full retraining on retained data. In addition, membership-inference results remain close to random-guessing behavior across most configurations, indicating limited privacy leakage after unlearning. Although small positive deviations appear under the most aggressive deletion settings, overall privacy behavior remains stable across practical operating regimes.

This study has several limitations. The evaluation used a 2{,}000-sample behavioral ransomware dataset, relied on a Q-margin-based MIA proxy rather than stronger adaptive attacks, and did not assess continual adaptation on streaming data. Future studies should therefore examine larger and more heterogeneous datasets, stronger privacy-auditing methods, and the interaction between repeated deletions and RL policy stability under more realistic deployment conditions.

From an information security perspective, the proposed framework addresses an important compliance gap for organizations using ML-based ransomware detection by enabling affected-shard retraining instead of full model rebuilding when behavioral telemetry must be removed. Its retraining efficiency (typically 5--30\,s versus 80--330\,s for full retraining), together with oracle verification and membership-inference auditing, supports compliance-oriented assessment without materially disrupting detection availability.

Overall, the results demonstrate that the DDQN--SISA framework enables ransomware detection with auditable machine unlearning, combining strong detection performance, selective data removal, and practical retraining efficiency in a privacy-compliant security setting.

\section{Conclusion and Future Work}

This study presents a privacy-compliant ransomware detection and unlearning framework that integrates DDQN-based detection with multi-shard SISA retraining. Unlike conventional ransomware detectors that assume fixed training data, the proposed framework supports selective shard-level retraining under deletion requests while preserving strong detection performance. The results show that selective retraining substantially reduces computational cost relative to full retraining, with unlearning time scaling with affected shards rather than total dataset size. Oracle-based evaluation further indicates that deleted samples no longer materially influence the retrained model, whereas membership-inference analysis shows limited distinguishability of removed samples after unlearning. Future work will evaluate the framework on larger and more heterogeneous datasets, investigate shard-allocation strategies at greater scale, and assess stronger adversarial privacy attacks and formal privacy guarantees. Additional studies are also needed to examine distributed deployment settings and the stability of repeated deletion operations in more realistic security environments. Overall, the findings demonstrate the promise of combining reinforcement learning and machine unlearning for auditable and privacy-compliant ransomware detection.

\printcredits

\bibliographystyle{cas-model2-names}

\bibliography{references}



\end{document}